\documentclass[useAMS,usenatbib]{mn2e}
\usepackage{amsmath}
\usepackage{txfonts}
\usepackage{graphics}
\usepackage{times}

\newcommand{\bx}{{\bf x}}
\newcommand{\bk}{{\bf k}}
\newcommand{\bq}{{\bf q}}
\newcommand{\bp}{{\bf p}}
\newcommand{\bP}{{\bf\Psi}}
\newcommand{\bs}{{\bf s}}

\begin{document}


\title[Galaxies and 21cm]{Combining galaxy and 21cm surveys}
\author[]{
    J.D. Cohn${}^{1}$,
    Martin White${}^{2,3}$,
    Tzu-Ching Chang${}^{4}$,
    Gil Holder${}^{5}$
    \newauthor
    Nikhil Padmanabhan${}^{6}$,
    Olivier Dor\'e${}^{7,8}$\\
    ${}^1$ Space Sciences Laboratory and Theoretical Astrophysics Center,
    University of California, Berkeley, CA 94720\\
    ${}^2$ Departments of Physics and Astronomy, University of California,
    Berkeley, CA 94720, USA \\
    ${}^3$ Lawrence Berkeley National Laboratory, 1 Cyclotron Road,
    Berkeley, CA 94720, USA \\
    ${}^4$ Academia Sinica Institute of Astronomy and Astrophysics,
   11F of ASMAB, AS/NTU, 1 Roosevelt Rd Sec. 4, Taipei 10617, Taiwan \\
    ${}^5$ Department of Physics, McGill University, 3600 rue University,
    Montreal, QC, H3A 2T8, Canada \\
    ${}^6$ Department of Physics, Yale University, New Haven, CT 06511, USA \\
    ${}^7$ Caltech M/C 350-17, Pasadena, CA 91125, USA\\
    ${}^8$ Jet Propulsion Laboratory, 4800 Oak Grove Drive, Pasadena, CA, USA
}

\pagerange{\pageref{firstpage}--\pageref{lastpage}}

\maketitle

\label{firstpage}

\begin{abstract}
Acoustic waves traveling through the early Universe imprint a characteristic
scale in the clustering of galaxies, QSOs and inter-galactic gas.
This scale can be used as a standard ruler to map the expansion history of
the Universe, a technique known as Baryon Acoustic Oscillations (BAO).
BAO offer a high-precision, low-systematics means of constraining our
cosmological model.
The statistical power of BAO measurements can be improved if the `smearing'
of the acoustic feature by non-linear structure formation is undone in a
process known as reconstruction.
In this paper we use low-order Lagrangian perturbation theory to study the
ability of $21\,$cm experiments to perform reconstruction and how augmenting
these surveys with galaxy redshift surveys at relatively low number densities
can improve performance.
We find that the critical number density which must be achieved in order
to benefit $21\,$cm surveys is set by the linear theory power spectrum near
its peak, and corresponds to densities achievable by upcoming surveys of
emission line galaxies such as eBOSS and DESI.
As part of this work we analyze reconstruction within the framework of
Lagrangian perturbation theory with local Lagrangian bias, redshift-space
distortions, ${\bf k}$-dependent noise and anisotropic filtering schemes.
\end{abstract}



\section{Introduction}
\label{sec:intro}

In the last decade it has been realized that the large-scale structure
in the Universe can be used as a tool for measuring its expansion history
with high accuracy and low systematics.
One of the premier methods for measuring the distance-scale and expansion
rate uses the baryon acoustic oscillation (BAO) `feature' as a calibrated,
standard ruler \citep[see][for a review]{PDG}.
Non-linear evolution of the large-scale structure in the Universe damps
the acoustic oscillations in the power spectrum at late times (as has been
extensively discussed in the literature,
e.g.,~\citealt{Bha96,TayHam96,MeiWhiPea99,CroSco08,PadWhi09,
McCSza12,TasZal12a,Sch15}).
The modes responsible for the broadening of the peak are of quite long
wavelength \citep{ESW07} and, as pointed out by \citet{Eis07}, these modes
are also generally well measured by a survey aiming to do BAO.
Thus the impact of the non-linear evolution can be modeled and reduced by
a process known as reconstruction \citep{Eis07,Pad12}.
Reconstruction greatly improves fits to the distance scale using the
BAO feature \citep{Seo10,And14}.

Traditionally BAO have been measured in galaxy surveys
\citep[e.g.,][for the most recent detections]{And14},
or in the intergalactic medium
\citep{Bus13,Slo13,Kir13,Del15}
either directly or in cross-correlation with QSOs
\citep{Fon14}.
In recent years technological advances have made it feasible to use
$21\,$cm surveys to measure large-scale structure and, in princple, the
BAO scale at redshifts $z\sim 1-2$.
In advance of a detection a wide variety of technologies are being
investigated, ranging from large arrays of dishes
(e.g.,~BAORadio\footnote{\citet{Ans12}}, HIRAX,
SKA1-MID\footnote{http://www.skatelescope.org}),
to large single dishes
(e.g.,~FAST\footnote{\citet{Nan11}}),
to arrays of antenna tiles
(e.g.,~BAOBAB\footnote{\citet{Pob13}}),
to arrays in the focal plane of a large reflector
(e.g.,~GBT-HIM,
Parkes,
BINGO\footnote{http://www.jb.man.ac.uk/research/BINGO}),
to arrays of cylindrical reflectors
(e.g.,~CHIME\footnote{http://chime.phas.ubc.ca},
ORT\footnote{\citet{AliBha14}},
Tianlai\footnote{http://tianlai.bao.ac.cn}).
Each of these approaches has its advantages and difficulties.
One difficulty that they all share in using redshifted $21\,$cm emission
as a cosmological probe is that the signal is dwarfed by
foreground\footnote{The foregrounds have been best studied in the context
of $21\,$cm studies of the epoch of reionization, i.e.,~at lower frequencies
than we consider.  However the signal and foregrounds scale in a similar
manner with frequency so that many of the results carry over with minimal
modification -- see e.g.,~\citet{Pob15} for a recent discussion.}
contamination \citep{Fur06,Cha10}.
This renders many longer-wavelength modes in such experiments unusable, and
this can have a significant impact on the ability of such experiments to
measure BAO \citep[see][for further discussion]{SeoHir15}.

The literature contains very different estimates for the ultimate impact
of foregrounds, depending upon assumptions about how well one can model the
instrument. Some authors claim that foregrounds can be removed down to
$k\simeq 0.02\,h\,{\rm Mpc}^{-1}$ \citep{Sha14,Sha15}
while others claim that all line-of-sight modes with
$k_\parallel<0.1\,h\,{\rm Mpc}^{-1}$ and modes with $k_\parallel<0.6\,k$
would be significantly contaminated
(e.g.,~\citealt{LiuParTro14,Pob15}, building upon
\citealt{DatBowCar10,Ved12,Par12,Mor12}).
We shall consider a range of possibilities motivated by these investigations.

Not surprisingly, the loss of long-wavelength modes causes a significant
reduction in the BAO signal and, even more dramatically, in the ability to
perform reconstruction on such a survey.
We note that at high redshift and large scales we expect the $k$-modes to
be almost independent of one another, and thus there is no interpolation or
filtering scheme that can compensate for a lost mode.\footnote{See however \citet{Zhu15}, which appeared as we were finishing this paper, for a method using the effect of long wavelength modes on short wavelength modes.}
If the $21\,$cm survey is unable to measure a given mode of the density field
it must be supplied by other means.
In this paper we investigate whether a very sparse tracer of the density
field, such as QSOs or emission line galaxies (ELGs), can be used to
recover some of the missing, large-scale modes and improve reconstruction.

Specifically we investigate how reconstruction is affected by missing modes
using low-order Lagrangian perturbation theory.
We extend this theory, in the context of reconstruction, to include
filtering and missing modes
\citep[see also][for similar topics]{Seo15} and anisotropic noise.
Missing modes can be thought of as modes with infinite noise.
A second sample can be used to ``fill in'' the modes missed by a $21\,$cm
survey, so that the noise in that region of $k$-space is set by the properties
of the second sample.
To be concrete, we consider QSOs and ELGs as tracers of the high-$z$,
large-scale density field, since surveys covering large areas of sky with
spectroscopic redshifts of objects in the appropriate redshift range
are in progress.  As an example, the
eBOSS\footnote{http://www.sdss.org/surveys/eboss} survey \citep{Daw15}
will obtain spectroscopic redshifts for more than 500,000 QSOs with
$0.9<z<2.2$ over $7500\,{\rm deg}^2$ of sky \citep{Mye15},
significantly extending the existing samples in this redshift range.
It will also measure redshifts for $190,000$ ELGs in the range $0.7<z<1.1$
over 750-1500 deg${}^2$.
In the future DESI\footnote{http://desi.lbl.gov} will generate samples
of emission line galaxies (and QSOs) with even higher number density and
wider redshift coverage.

As might be expected, there is little gain in using tracers whose
shot-noise exceeds their clustering power on the scales relevant for
computing the displacements.  Which tracers are useful in this context
thus depends on the modes which a $21\,$cm survey is unable to access.
In its standard configuration, QSOs are shot-noise limited at all $k$
for the eBOSS surveys, though one could imagine similar surveys which
could go deeper.  Similarly the Ly$\alpha$ forest measured by BOSS and
eBOSS provides good sampling, but long-wavelength modes along the line
of sight can be contaminated by continuum modeling.
Perhaps the best choice is the eBOSS ELG survey near $z\simeq 1$ which
will provide a useful sample to augment reconstruction or, in the future,
the DESI sample where even the lower density (but higher bias), high-$z$
tail will be of use.

The outline of the paper is as follows.
In section \S\ref{sec:lagiso}, we use low-order Lagrangian perturbation
theory to characterize reconstruction in the presence of noise, and then
consider missing modes as modes with infinite noise.
To avoid modes with infinite noise from erasing all the effects of
reconstruction, we introduce a Wiener filter into the reconstruction
scheme.
In section \S\ref{sec:anisotropic}, we generalize our treatment to include
anisotropic noise (such as the ``wedge'' for $21\,$cm experiments; \citealt{DatBowCar10}) and
redshift space distortions and bias, and then show that the addition of
even sparsely sampled objects to data where modes were previously missing can
improve reconstruction.  We show how this depends upon the noise of the
added field and the geometry of the ``wedge''.
We conclude in section \S\ref{sec:discussion}.
A discussion of Eulerian perturbation theory is provided in an appendix,
as it is a particularly physical way of viewing reconstruction and seeing
how the missing modes limit the effects of reconstruction.

\section{Reconstruction with noise}
\label{sec:lagiso}

Lagrangian perturbation theory has proven particularly useful as an
approximate, analytic model of reconstruction
\citep{PadWhiCoh09,NohWhiPad09,TasZal12b,Whi15,Seo15}.
In this section we review the Lagrangian framework, the reduction in
signal-to-noise ratio that arises from non-linear structure formation and
how to model reconstruction within this formalism.
We begin with reconstruction for an isotropic system, to set notation
and identify a few key features when noise is included.
The generalization to include anisotropic noise and redshift-space
distortions is presented in section \S\ref{sec:anisotropic}.

\subsection{Review: Peak broadening in Lagrangian Perturbation Theory}

The modification of the power spectrum by non-linear evolution
can be seen in the Lagrangian approach as follows
\citep[e.g.,][]{Mat08,PadWhiCoh09}.
If we denote a particle's initial, Lagrangian, position by $\bq$ and its
final, Eulerian, position by $\bx$ then the Lagrangian displacement is
defined through $\bx=\bq+\bP(\bq)$.  Assuming a local, Lagrangian bias
defined by $F[\delta_L(\bq)]$ the density is \citep{Mat08}
\begin{equation}
  1+\delta(\bx) = \int d^3q\ F[\delta_L(\bq)]\,\delta^{(D)}(\bx-\bq-\bP)
\end{equation}
with
\begin{eqnarray}
  \delta(\bk) &=& \int d^3x\ e^{-i \bk \cdot \bx}\, \delta(\bx) \nonumber \\
  &=& \int d^3q\,e^{-i \bk\cdot \bq}
  \left(\int\frac{d\lambda}{2\pi}\ F(\lambda) e^{-i\bk\cdot\bP(\bq)-i\lambda\delta_L} -1\right) \; .
\label{eqn:deltak}
\end{eqnarray} 
An unbiased tracer of the density field has
$F[\lambda]\propto\delta^{(D)}(\lambda)$,
while a tracer with linear bias $b$ has the average of $F'$ over the Gaussian distribution
of $\delta_L$ equal to $b-1$ \citep{Mat08}.
To leading order in the linear density contrast, $\delta_L$,
\begin{equation}
  \bP(k) = i \frac{\bf k}{k^2} \delta_L ({\bf k})
\end{equation}
\citep{Zel70}.
Using the cumulant theorem and the fact that $\delta_L$, and hence $\bP$,
is Gaussian one obtains [see Appendix \ref{app:derivation}] a damping of
the oscillations
\citep[e.g.,][]{Bha96,TayHam96,MeiWhiPea99,ESW07,CroSco08,Mat08}
\begin{equation}
  P_{\rm nl}(k) = \mathcal{D}(k)\, b^2 P_L(k) + \cdots
  \quad {\rm with}\quad
  \mathcal{D}(k)=e^{-k^2\Sigma^2/2}
\label{eqn:pkform}
\end{equation}
(We have assumed a scale-independent bias as appropriate for large scales
and $P_L$ is the linear dark matter power spectrum.)
The damping of the linear power spectrum (or equivalently the smoothing of
the correlation function) reduces the contrast of the feature and
the precision with which the size of ruler may be measured.
The damping scale is set by the mean-squared Zel'dovich displacement
of particles,
\begin{equation}
  \Sigma^2 = \frac{1}{3\pi^2} \int dp\ P_L(p) \qquad .
\label{eqn:sigmal}
\end{equation}
As we shall see, reconstruction decreases $\Sigma^2$ (and also undoes the
shift of the peak position from its linear value --
see \citealt{PadWhiCoh09,SheZal12} for further discussion).
We will be studying the effects of reconstruction by considering changes
to $\mathcal{D}(k)$.

\subsection{Review: Reconstruction algorithm}
\label{sec:recalg}

Reconstruction `undoes' the effects of non-linearity using the measured
large-scale density field to infer the shifts that galaxies have undergone
due to gravitational instability.
The algorithm devised by \citet{Eis07} consists of the following steps:
\begin{enumerate}
\item Smooth the halo, galaxy or 21cm density field with a kernel $W$
(see below) to filter out small scale (high $k$) modes, which are difficult
to model.  Divide the amplitude of the overdensity by an estimate of the
large-scale bias, $b$, to obtain a proxy for the overdensity field:
$\delta(\mathbf{x})$.
\item Compute the shift, $\mathbf{s}$, from the smoothed density field in
redshift space using the Zeldovich approximation
(this field obeys $\mathbf{\nabla}\cdot\mathbf{R}\mathbf{s}=-\delta$ with
 $R_{ij}=\delta_{ij}+(f/b)\hat{z}_i\hat{z}_j$).
Once $\mathbf{s}$ is obtained, multiply the line-of-sight component
by $1+f$ to approximately account for redshift-space distortions (see
below for further discussion).
\item Move the galaxies by $\mathbf{s}$ and compute the ``displaced''
density field, $\delta_d$.
\item Shift an initially spatially uniform distribution of particles by
$\mathbf{s}$ to form the ``shifted'' density field, $\delta_s$.
\item The reconstructed density field is defined as
$\delta_r\equiv \delta_d-\delta_s$ with power spectrum
$P_{\rm rec}(k)\propto \langle \left| \delta_r^2\right|\rangle$.
\end{enumerate}
Following \citet{Eis07} we use a Gaussian smoothing of scale $R$ for $W$,
specifically $W(k)=\exp[-(kR)^2/2]$.  Our $R$ is canonically defined for
a Gaussian smoothing, but alternative definitions of $R$ exist in the
literature\footnote{For example, \citet{PadWhiCoh09} use
  $W(k)=\exp[-(kR)^2/4]$ in a similar analytic approach, while
  \citet{Bur14, Var14} use the definition here.}, so care must be taken in comparisons.
We take $R=10\,h^{-1}$Mpc unless otherwise noted.
Throughout we shall assume that the fiducial cosmology, bias and
$f=d\ln D/d\ln a\simeq\Omega_m^{0.55}$ are properly known during
reconstruction.
Various tests of the \citet{Eis07} reconstruction algorithm and
sensitivity to parameter choices have been performed in the literature.
We refer the reader to \citet{Seo10,Pad12,Xu13,Bur14,Toj14,Var14,Seo15}
which also contain useful details on the specific implementations.
Note that we have chosen to perform `anisotropic reconstruction', in which
the shifted and displaced fields both include the factor of $1+f$ in the
line-of-sight direction as Lagrangian perturbation theory seems to model
this algorithm better \citep{Whi15}.
Our implementation of anisotropic reconstruction follows \citet{Whi15} and
differs slightly from that in \citet{Seo15}.
We correct for redshift-space distortions (to linear order) in defining
$\mathbf{s}$ in terms of the observed density field but include the factor
of $1+f$ in the line-of-sight component for both the displaced and shifted
fields.  We shall make further comparison with \citet{Seo15} later.

With these definitions the displaced field is\footnote{See \citet{Whi15},
\S\,3.1, for a discussion of the use of $\bs(\bx)$ vs.~$\bs(\bq)$.}
\begin{equation}
 \delta_d(k) = \int d^3q\,e^{-i\bk\cdot\bq}
 \left(\int\frac{d\lambda}{2\pi}\ F(\lambda)
 e^{-i\bk\cdot(\bP(\bq)+\bs(\bq))-i\lambda\delta_L(\bq)} -1\right) \; , 
\end{equation}
while the ``shifted'' density field is
\begin{equation}
  \delta_s(k) = \int d^3q\ e^{-i \bk\cdot \bq}
  (e^{-i \bk\cdot \bs(\bq)} -1) \; .
\end{equation}
and hence
\begin{equation}
  \delta_r(k) = \int d^3q e^{-i\bk\cdot\bq} e^{-i\bk\cdot\bs(\bq)}
  \left(\int\frac{d\lambda}{2\pi}\ F(\lambda)
  e^{-i\bk\cdot\bP(\bq)-i\lambda\delta_L(\bq)}-1\right)
  \; , \\
\label{eqn:reconfield}
\end{equation}
The reconstructed power spectrum is
$P(k)\propto \langle \left| \delta_r^2\right|\rangle$, which
can be evaluated through use of the cumulant theorem.

\begin{figure}
\begin{center}
\resizebox{\columnwidth}{!}{\includegraphics{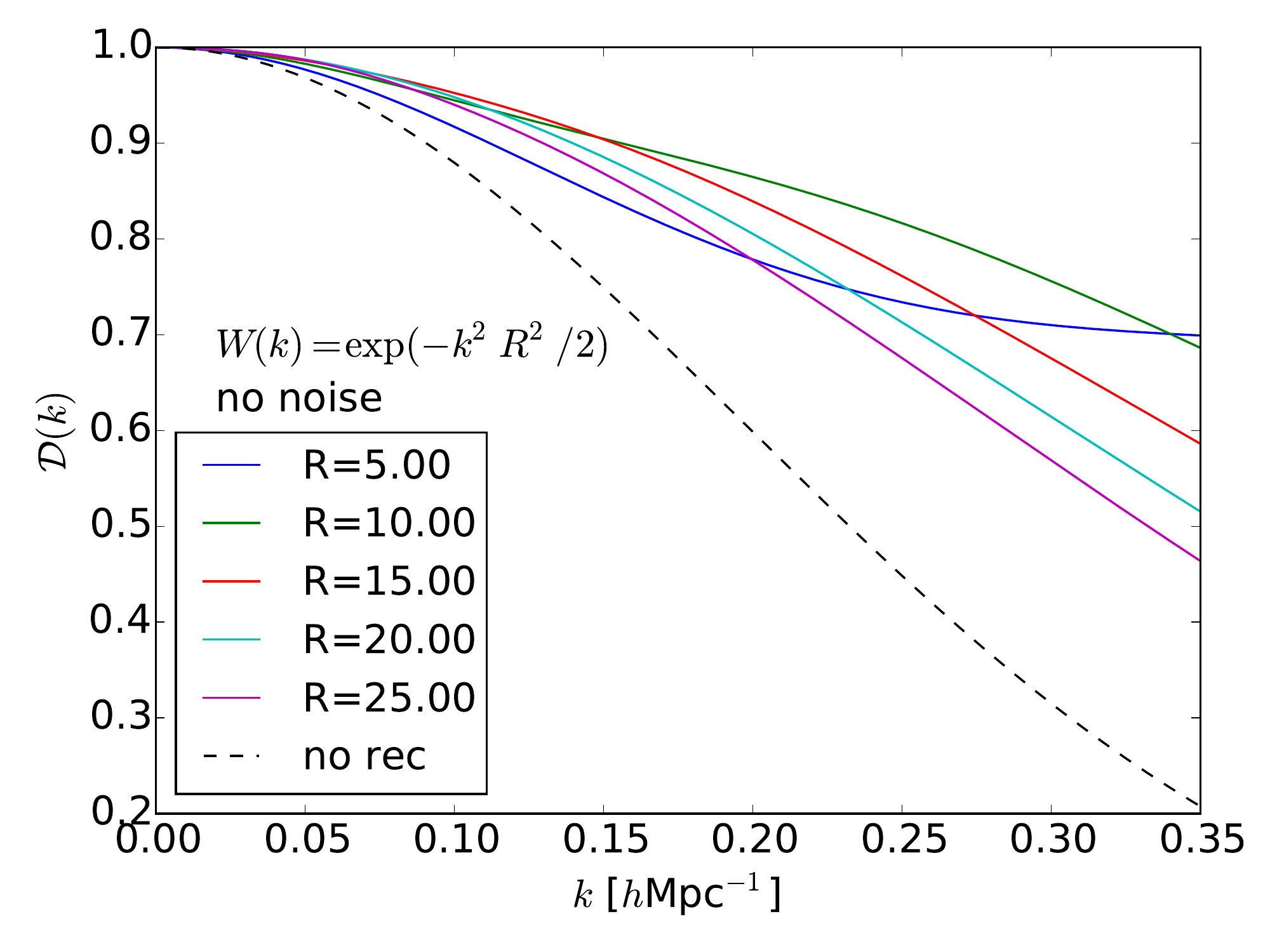}}
\resizebox{\columnwidth}{!}{\includegraphics{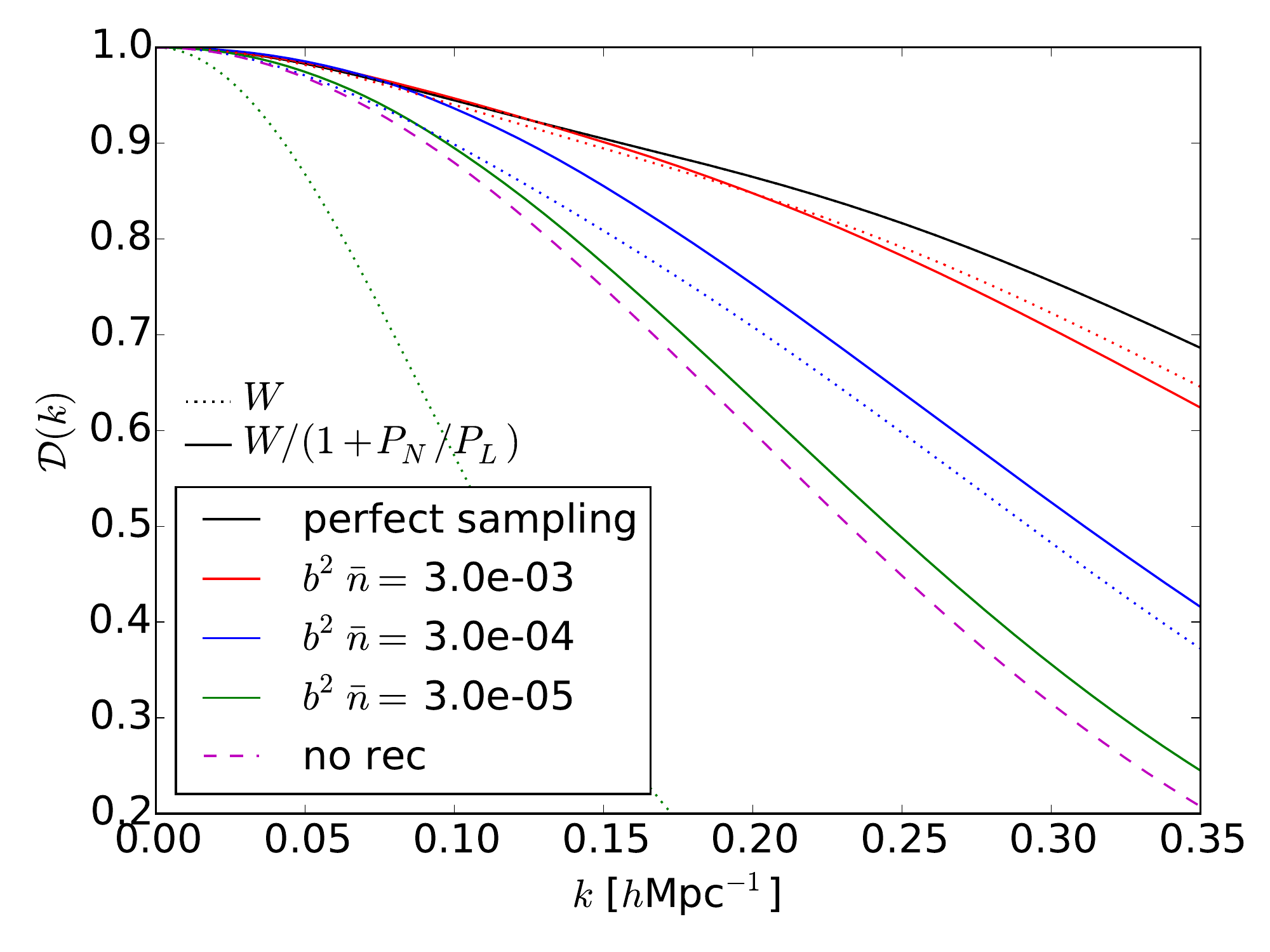}}
\end{center} 
\caption{ The leading coefficient $\mathcal{D}(k)$ of the
  reconstructed power spectrum, $P_{\rm rec}(k)={\mathcal
    D}(k)P_L(k)+\dots$, for different smoothings and filters.  The
  larger $\mathcal{D}(k)$, the better the reconstruction for the mode
  $k$.  
Top: the change in ${\mathcal D}(k)$ with changes in the
  Gaussian smoothing scale $R$.  For filtering scale below $\sim
  10\,h^{-1}$Mpc, $\mathcal{D}(k)$ is larger at large $k$ and smaller
  at low $k$ than for $R>10\,h^{-1}$Mpc. 
 Bottom: Holding the
  smoothing $R=10\,h^{-1} Mpc$, ${\mathcal D}(k)$ for different
  filters and four different choices of isotropic Poisson noise,
  corresponding to 4 densities.  [Densities are quoted in units of
  $h^3\,{\rm Mpc}^{-3}$.] 
Similar to changing the smoothing scale $R$
  above, increasing the noise level (decreasing the density $b^2
  \bar{n}$) can improve reconstruction [increase ${\mathcal D}(k)$]
  for some filters and values of $k$ and degrade reconstruction for
  others.  This can be seen in Fig.~\ref{fig:missingmodes} below.
  Also shown is the unreconstructed signal (magenta dashed line) in
  the absence of noise.  Here and hereon, redshift $z=1$ is assumed.}
\label{fig:scalediff}
\end{figure}

\subsection{Reconstruction with noise}

In current galaxy redshift surveys targeted at BAO, the number density
of tracers is such that noise is relatively insignificant.  However this
need not be the case in general.
When noise is present in the observed field, it will also
contribute to the shifted field \citep{Whi10}.
Assuming that the signal approaches linearity on large scales, for
sufficiently large smoothing the general expression for the shifted
field obeys
\begin{equation}
\begin{array}{ll}
  \bs(\bk) 
  & \simeq -i(\bk/k^2)\ \mathcal{S}(\bk)
  \ \left[ \delta_L(\bk) + \delta_N(\bk) \right] \; .
\end{array}
\label{eqn:shiftfield}
\end{equation}
where we have written the noise contribution $\delta_N({\bf k})$
explicitly.\footnote{Note that there is an implicit requirement on the
smoothing $\mathcal{S}({\bf k})$: when $\mathcal{S}$ acts on the observed
field it must suppress the non-linear power.}

In Eq.~\ref{eqn:shiftfield} the weight, $\mathcal{S}(k)$, applied to the
density field is a combination of the usual Gaussian smoothing kernel and
a noise-suppression factor,
\begin{equation}
\mathcal{S}(\bk)=f_N\left(P_L(\bk),P_N(\bk)\right) \,W(k) \; .
\label{eq:sfildef}
\end{equation}
Here $P_L$ and $P_N$ are inputs to the algorithm, taken to be theoretical,
linear power spectra estimates for the signal and noise respectively.
We shall assume these are the `true' signal and noise spectra, since the
procedure can be iterated.

We have not attempted to optimize $f_N$.  Instead we have chosen it to be a
Wiener filter which acts to suppress noise in $\bs$ from modes where $P_N$
is as large (or larger) than the cosmological signal.
\begin{equation}
  f_N \equiv \frac{P_L(\bk)}{P_L(\bk)+P_N(\bk)}
  = \frac{1}{1+[P_N({\bf k})/P_L({\bf k})]} \; .
\label{eq:wien}
\end{equation}
If such a factor $f_N$ is not included then a very noisy density field will
lead to almost random ``shifts'' which wash out the structure rather than
reconstructing it. 
A Wiener filter was also suggested by \citet[][Eq.~17]{SeoHir15}, though
they in practice used a Gaussian with a modified smoothing instead.

\subsection{Isotropic example}

The effects of the noise $P_N$ and the
filtering $f_N$ can be seen clearly in a simplified isotropic example, where
$P_N({\bf k})= P_N(k)$, redshift space distortions are neglected,
and the bias $b=1$.
In this limit, the leading terms of the reconstructed power spectrum are
$P_{\rm rec}({\bf k})=\mathcal{D}({\bf k})P_L(k) + \dots$
with\footnote{The propagator is $\sqrt{\mathcal{D}}$.}
\begin{equation}
  \mathcal{D}(k) = e^{-(1/2)k^2\Sigma_{ss}^2}
  \left[\mathcal{S}(k)+ \left(1-\mathcal{S}(k)\right)
  e^{-(1/4)k^2(\Sigma^2 - 2\Sigma^2_s)}\right]^2 \; .
\label{eq:noiseiso}
\end{equation} 
Here
\begin{eqnarray}
  \Sigma_{ss}^2 &=& \frac{1}{3\pi^2} \int dp\ \mathcal{S}(p)^2
  \left[P_L(p)+P_N(p)\right] \nonumber \\
  \Sigma_{s}^2  &=& \frac{1}{3\pi^2} \int dp\ \mathcal{S}(p) P_L(p)
\label{eq:sigsiso}
\end{eqnarray}
and $\Sigma^2$ is the unreconstructed damping (Eq.~\ref{eqn:sigmal}).
This is equivalent to, but written slightly differently than, previous
results (see later).  This form isolates both the damping $\Sigma_{ss}^2$
remaining after reconstruction (which also contains the full effect of
the noise in this approximation) and the damping $\Sigma^2$ in the absence
of reconstruction.
With no filter (i.e.,~$f_N=1$), increasing the noise clearly increases
$\Sigma_{ss}^2$, increasing the overall damping of the reconstructed
linear power spectrum.\footnote{Note that the presence of noise means
that the shifted field also has a noise component
$\delta_r=\delta_d-\delta_s + \delta_N$ so
that to leading order there is a term proportional to noise $P(k) =
{\mathcal D}({\bf k}) P_L(k) + {\mathcal D}_N(({\bf k})
P_N({\bf k})+ \dots$, where ${\mathcal D}_N(k) = 1 + \dots$,
i.e., the leading contribution to the noise is not damped by reconstruction.}

Taking $f_N$ to be the Wiener filter, $1/(1+P_N/P_L)$, the contributions
$\mathcal{S}^2[P_L + P_N]$ in the damping scale in Eq.~\ref{eq:sigsiso}
go to zero when $P_N(\bk)$ is large
($\mathcal{S}P_L $ also tends to zero in this limit).
The very noisy modes thus do not contribute significantly to $\Sigma_{ss}^2$
or to $\Sigma_s^2$, i.e.,~to reconstruction.
Specifically, $\Sigma_{ss}^2$ no longer increases with increasing noise as
it would if no filter were applied.
In the examples below we shall take the noise to be Poisson shot noise,
$P_N=b^{-2}\bar{n}^{-1}$, for tracers with number density $\bar{n}$ and
linear bias $b$.  In fact, we shall typically quote the noise levels in terms
of an effective number density.

A filter can not only ensure that noisy modes do not destroy reconstruction
for all modes; it also modifies reconstruction for the remaining modes, as
does the form and level of noise.  These effects can be similar to those due
to changing the Gaussian smoothing scale $R$.
(Studies changing $R$ include \citealt{Pad12, Bur14,Seo15}; as mentioned
earlier \citet{SeoHir15} trade a change in $R$ for a filter.)
Altering the filter, Gaussian smoothing or noise level does not necessarily
have a uniform effect on the reconstruction of different modes: a given
change may increase damping for some modes and decrease it for others.
In Fig.~\ref{fig:scalediff} we show the change in damping, $\mathcal{D}(k)$,
when  changing the smoothing $R$ (top panel, fixing $f_N=1$ by setting
$P_N=0$)  as well 
as changing the noise
level at fixed $R$ (bottom panel).  In the latter case we show the results for two filters:
$f_N=1$ and $f_N=1/(1+P_N/P_L)$.
Modes at low $k$ and high $k$ have different trends in how well they are
reconstructed as the smoothing increases or the filter changes.
In particular with a noise dependent filter and smaller $R$
$(R< 10\,h^{-1}$Mpc, not shown) increasing noise can increase
$\mathcal{D}$ for some low values of $k$.    
To summarize, the combination of noise, filter and smoothing affects
different $k$-modes differently, and increasing noise can sometimes
improve reconstruction of a specific $k$-mode.
This suggests that a series of filters, each optimized to a specific
window and with an optmized shape, could improve reconstruction over
what is possible with a single smoothing scale.  
We shall leave such investigation to a future paper.
Henceforth we fix $f_N = 1/(1+P_N({\bf k})/P_L({\bf k}))$.

Now we are in a position to describe the effects of missing modes,
i.e.,~modes which are not observed, on reconstruction.
First of all, these modes are not present for measuring the power
spectrum, which decreases the number of modes which can be averaged
over and thus increases the error in a given $k$-bin [perhaps to infinity].
Their absence also weakens the ability to reconstruct the rest of the modes.
By treating missing modes as modes with infinite noise
in Eq.~\ref{eq:noiseiso}, they can be seen to contribute to the original
broadening ($\Sigma^2$ is unchanged) but not to the reconstructed factors
$\Sigma_{ss}^2$ and $\Sigma_s^2$.  Before we explore this further we
generalize our treatment to include anisotropic reconstruction and
redshift-space distortions.

\section{Anisotropic reconstruction}
\label{sec:anisotropic}

\subsection{Formalism}

Including redshift-space distortions in our theory, the displacements $\bP$
and $\bs$ are increased in the line-of-sight direction by a factor
$1+f$ as noted in section \S\ref{sec:recalg}.
The undamped signal is modified from $b^2\,P_L$ to $(b+f\mu^2)^2P_L$,
where we have again assumed linear bias $b$.
In addition, the noise power spectra appearing in the expressions below
are to be interpreted as the raw noise power divided by $[1+\beta\mu^2]^2$,
\begin{equation}
  P_N ({\bf k}) \rightarrow \frac{P_N({\bf k})}{[1+\beta\mu^2]^2} \; ,
\end{equation}
since we ``remove'' redshift-space distortions when first computing
$\bs$ from our smoothed density field. 
This additional factor also propagates into the Wiener filter expression
and thus $\mathcal{S}({\bf k})$.
The noise power, $P_N$, may depend on the direction of $\bk$ beyond that
given by the $1+\beta\mu^2$ term; for example the $21\,$cm wedge corresponds
to direction dependent missing modes (i.e.,~modes with very large noise).

The reconstructed power spectrum thus reads
\begin{equation}
  P_{\rm rec}(k,\mu) = (b+f \mu^2)^2 \mathcal{D}({\bf k}) P_L(k) +
  \cdots 
\end{equation}
with damping
\begin{eqnarray}
  \mathcal{D}({\bf k})& =& 
  e^{-(1/2)\{k_\perp^2\Sigma_{ss,\perp}^2+
             k_\parallel^2\Sigma_{ss,\parallel}^2\}}
  \left[\frac{1+f\mu^2}{b+f\mu^2}\, \mathcal{S}(\bk) + \right.  \\
  && \left.  \left(1-\frac{1+f\mu^2}{b+f\mu^2}\,\mathcal{S}(\bk)\right)
  e^{-(1/4)\{(k_\perp^2+(1+f)^2k_\parallel^2)\Sigma^2
  -2k_\perp^2\Sigma_{s,\perp}^2- 2 k_\parallel^2\Sigma_{s,\parallel}^2\}}
  \right]^2 \; . \nonumber \\
\label{eqn:dampmissing2}
\end{eqnarray}

The damping terms again depend upon the linear power spectrum and the
(now possibly anisotropic) noise, and are given by:\footnote{The
 generalization to noise which depends on $k_x$ and $k_y$ separately
 is straightforward.}
\begin{eqnarray}
  \Sigma_{ss,\parallel}^2 &=& 2(1+f)^2\int \frac{d^3p}{(2\pi)^3}
  \frac{ p_\parallel^2}{p^4}\  \mathcal{S}^2(\bp) \left[ P_L(\bp)+P_N(\bp) \right]
  \nonumber \\
  \Sigma_{ss,\perp}^{2} &=& \hphantom{2(1+f)^2}\int \frac{d^3p}{(2\pi)^3}
  \frac{p_\perp^2}{p^4}\  \mathcal{S}^2(\bp) \left[ P_L(\bp) + P_N(\bp) \right]
  \nonumber \\
  \Sigma_{s,\parallel}^{2} &=& 2(1+f)^2\int\frac{d^3p}{(2\pi)^3}
  \frac{p_\parallel^2}{p^4}\ \mathcal{S}(\bp) P_L(\bp)  \nonumber \\
  \Sigma_{s,\perp}^{2} &=&\hphantom{2(1+f)^2}\int\frac{d^3p}{(2\pi)^3}
  \frac{p_\perp^2}{p^4}\  \mathcal{S}(\bp) P_L(\bp)  \, 
\end{eqnarray}

These expressions are equivalent to the more familiar forms
\begin{equation}
\begin{array}{lll}
  \mathcal{D}({\bf k}) &=&\mathcal{S}^2({\bf k})
  e^{-(1/2)(k_\perp^2\Sigma_{ss,\perp}^2+
  k_\parallel^2\Sigma_{ss,\parallel}^2)} \\
  &+&  [1-\mathcal{S}({\bf k})]^2
  e^{-(1/2)(k_\perp^2\Sigma_{dd,\perp}^2+
  k_\parallel^2 \Sigma_{dd,\parallel}^2)} \\
  &+& 2\mathcal{S}({\bf k})[1-\mathcal{S}({\bf k})]
  e^{-(1/2)(k_\perp^2\Sigma_{sd,\perp}^2+
  k_\parallel^2\Sigma_{sd,\parallel}^2)}\,\,.
\end{array}
\label{eqn:dampmissing}
\end{equation}
where we have taken $b=1$ and defined
\begin{eqnarray}
  \Sigma_{dd,\parallel}^{2} &=& 2(1+f)^2\int\frac{d^3p}{(2\pi)^3}
  \frac{p_\parallel^2}{p^4} \left\{ \left[1-\mathcal{S}(\bp)\right]^2
  P_L(\bp)+\mathcal{S}^2(\bp)P_N(\bp) \right\}  \nonumber \\
  \Sigma_{dd,\perp}^{2} &=& \hphantom{2(1+f)^2}\int\frac{d^3p}{(2\pi)^3}
  \frac{p_\perp^2}{p^4}
  \left\{ \left[1-\mathcal{S}(\bp)\right]^2 P_L(\bp) +
  \mathcal{S}^2(\bp)P_N(\bp) \right\} \, ,  \nonumber
\end{eqnarray}
with
$\Sigma^2_{sd,\perp}=(1/2)(\Sigma^2_{dd,\perp}+\Sigma^2_{ss,\perp})$
and $\Sigma^2_{sd,\parallel}$ defined similarly.
Note that for isotropic $\mathcal{S}$ and $P_N$ the parallel components
are simply $(1+f)^2$ times the perpendicular components and the coefficients
of the exponentials become
\begin{equation}
  \frac{k^2\Sigma^2}{2}\left\{(1-\mu^2) + (1+f)^2\mu^2\right\}
  = \frac{k^2\Sigma^2}{2}\left\{1+f(f+2)\mu^2\right\}
  \quad .
\end{equation}
In the appropriate limits these equations then agree with those in
\citet{Mat08,Whi10,CarReiWhi13,Seo15}.
Note that in the notation of the Appendix of \citet{Seo15} we have
chosen $\kappa=b+f\mu^2$ and $\lambda_d=\lambda_s=f$.
It is closest to their ``rec-iso'' case (in which they choose
$\lambda_s=0$).

\subsection{Example: $21\,$cm wedge}

\begin{figure}
\begin{center}
\resizebox{\columnwidth}{!}{\includegraphics{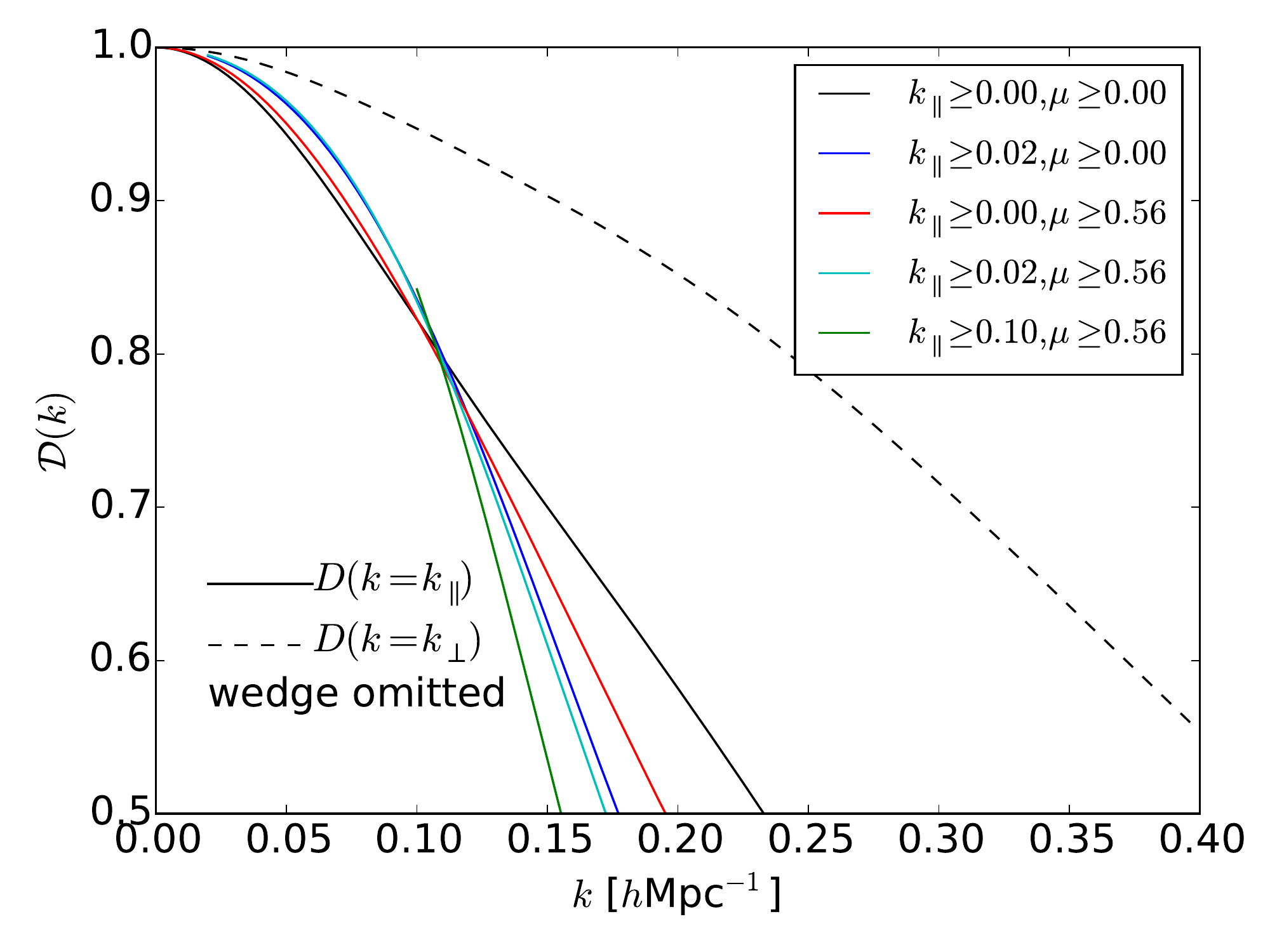}}
\resizebox{\columnwidth}{!}{\includegraphics{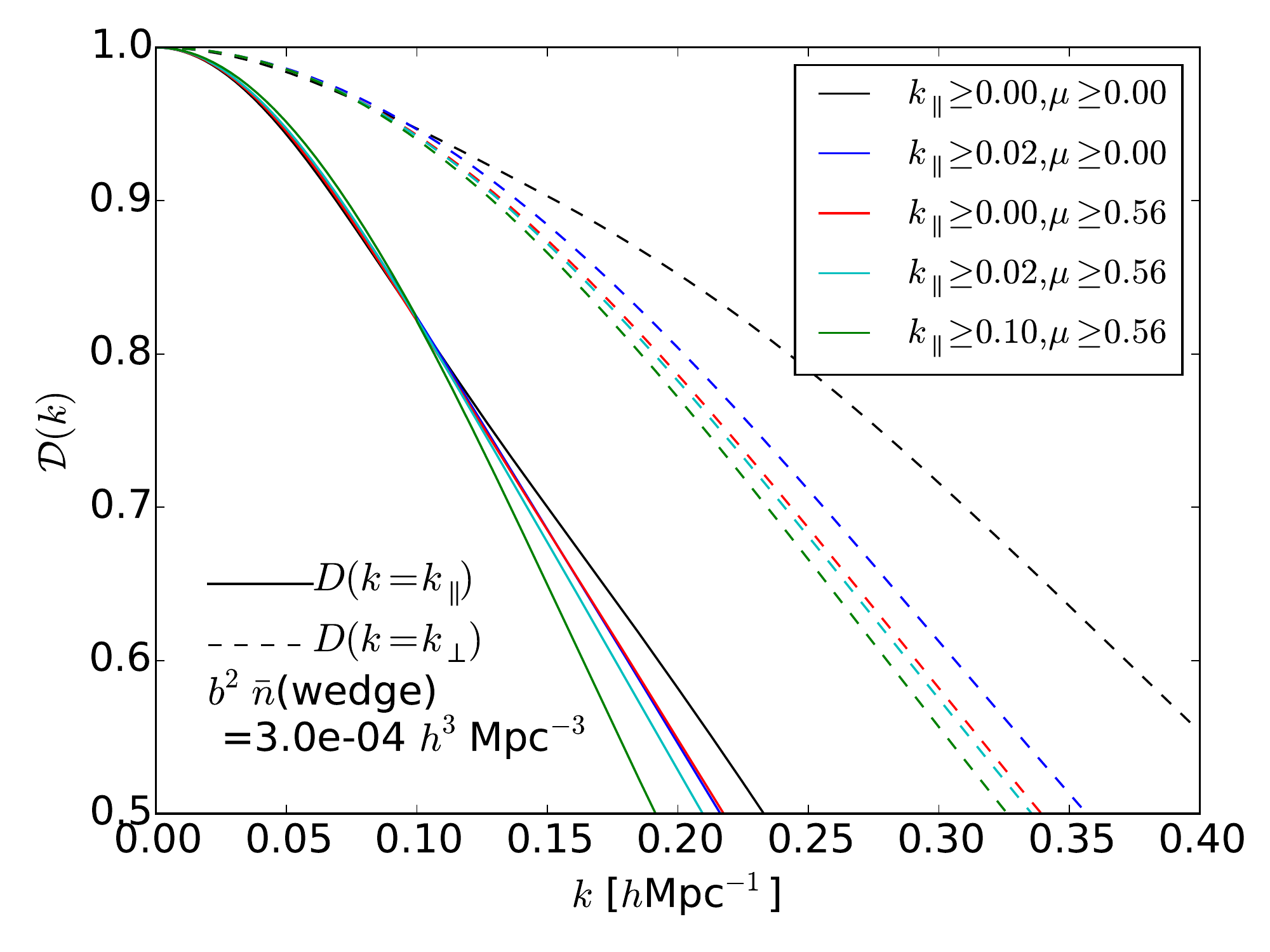}}
\end{center} 
\caption{The damping factor, $\mathcal{D}(k)$, for a survey where $21\,$cm
modes have $k_\parallel$ and $\mu=|k_\parallel|/k$ restricted to the ranges
shown.  The black lines
have no missing modes.  Colored lines include the
effect of the wedge with $\mu_{\rm min}=0.56$ and/or a cut in
$k_\parallel$;
the two lines for the constraints
$k_\parallel \ge 0.1 h Mpc^{-1}$, $\mu \ge 0, 0.56$ are degenerate.
The noise of the $21\,$cm modes, i.e., modes not in the wedge, is taken to be
equivalent to $b^2\,\bar{n}\simeq 3\times 10^{-3}\,h^3{\rm Mpc}^{-3}$
(see text). 
Two directions are shown: solid lines are along the line-of-sight
($k_\parallel=k$, $k_\perp=0$) and dashed lines are transverse
($k_\perp=k$, $k_\parallel=0$).
Top: $\mathcal{D}(k)$ for modes which are present (obeying
$k_\parallel$ and $\mu_{\rm min}$ cut), 
$P_N=\infty$ or $b^2 \bar{n}\to 0$ for missing modes.
Bottom: The same cuts in $k_\parallel$, $\mu$, as above, but replacing the
missing modes with an ELG survey with number density
$3\times 10^{-4}\,h^3\,{\rm Mpc}^{-3}$.  
Note in the top figure that with our approximations a small range of $k_\parallel$ is improved when
other modes are completely left out of reconstruction, an example of
increasing noise increasing $\mathcal{D}(k)$ and suggesting that a
different smoothing may help with better recovering those components.}
\label{fig:missingmodes}
\end{figure}

As our motivating example we consider a density field with a noise level
and different missing mode scenarios inspired by $21\,$cm experiments.
We will take as a representative $21\,$cm noise level that corresponding
to the forecast for a CHIME-like experiment in \citet{Cha08}, assuming that
thermal noise dominates \citep[see also][]{Seo10b,SeoHir15}.
The CHIME-like noise can be expressed in terms of an effective number density
of tracers (of unit bias)
\begin{eqnarray}
  \bar{n} &=& 2.5\times 10^{-3}\,h^3{\rm Mpc}^{-3}
  \left(\frac{1+z}{2.5}\right)^{3/2} \nonumber \\
  &\times&
  \left[\frac{\Omega_\Lambda(1+z)^{-3}+\Omega_m}{0.3}\right]^{-1/2}
  \left(\frac{\chi}{3.2\,h^{-1}{\rm Gpc}}\right)^{-2}
\label{eq:chimenoise}
\end{eqnarray}
With an eye to using ELG's as the example of a ``helper'' tracer we will
focus on redshift $z=1$.  In this case
$\bar{n}\simeq 3\times 10^{-3}\,h^3\,{\rm Mpc}^{-3}$.

For missing modes, we consider scenarios where we include only modes with
$|k_\parallel|>k_{\rm cut}$ which lie outside a ``wedge'',
$|k_\parallel|/k>\mu_{\rm min}$.
For definiteness, and because it covers the wide range of opinions in the
literature, we consider $k_{\rm cut}=0$, 0.02, $0.1\,h\,{\rm Mpc}^{-1}$.
For $\mu_{\rm min}$ the edge of the wedge is given by
\citep[e.g.,][and references therein]{Sha15,Pob15,SeoHir15}
\begin{equation}
  \frac{k_\parallel}{k_\perp} \equiv \mathcal{R}
  = \frac{\chi(z) H(z)}{c(1+z)}
  = \frac{E(z)}{1+z} \int_0^z \frac{dz'}{E(z')}
\end{equation}
where $\chi$ is the comoving distance to redshift $z$, $E(z)=H(z)/H_0$ and
we have assumed a spatially flat Universe.
The appropriate prefactor in front of $\mathcal{R}$ and how far into the wedge
it is possible to work is a matter of debate -- we shall take this value as
illustrative though higher values are possible.
If the wedge cannot be penetrated then all modes with
$|\mu|\equiv |k_\parallel|/k < \mu_{\rm min}$
\begin{equation}
  \mu_{\rm min} = \frac{\mathcal{R}}{\sqrt{1 + \mathcal{R}^2}}
\end{equation}
are lost.  For a $\Lambda$CDM model at $z=1$, and our choice of edge,
$\mu_{\rm min}\simeq 0.56$.
Thus slightly more than half of all modes would be lost to the wedge
at $z=1$.
A graphical representation of the modes which are lost (or kept) can be
seen in Fig.~\ref{fig:fullexample} below.

We show in the top panel of Fig.~\ref{fig:missingmodes} the change in $\mathcal{D}(k)$, for
$k=k_\parallel$ or $k=k_\perp$ when certain modes are omitted from the
reconstruction (the black solid and dashed lines are the comparison
case of no missing modes).
As already noted by \citet{SeoHir15}, the effect on reconstruction is
dramatic.
Missing modes thus significantly compromise the ability of a $21\,$cm
survey to measure the distance scale.

\subsection{Filling in the wedge}
\label{sec:combination}

\begin{figure}
\begin{center}
\resizebox{\columnwidth}{!}{\includegraphics{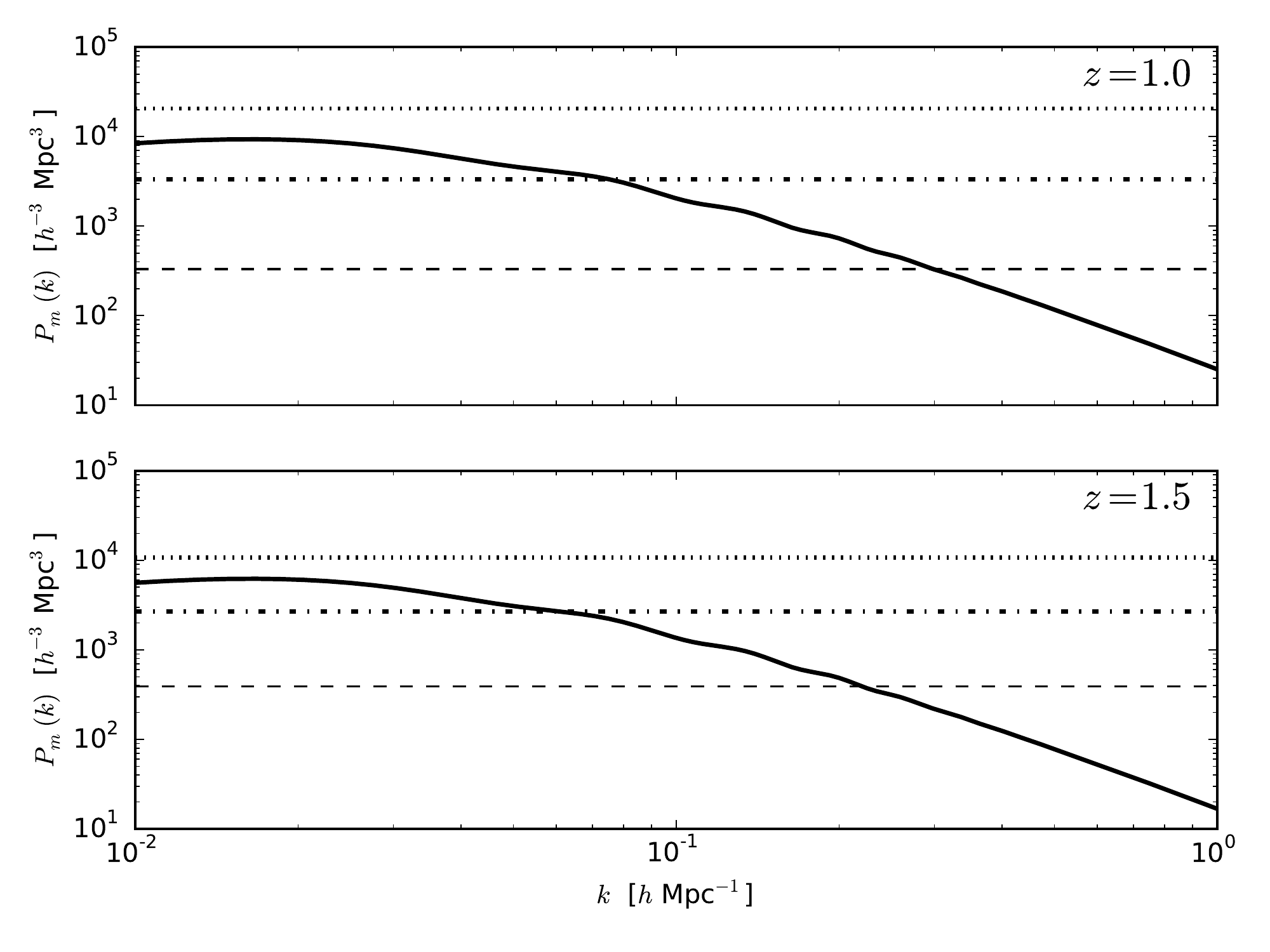}}
\end{center}
\caption{The matter power spectrum, at $z=1$ and 1.5, compared to the
QSO shot-noise ($1/b^2\bar{n}$; dotted line), the ELG shot-noise (dot-dashed)
and the noise expected for a $21\,$cm survey such as CHIME (dashed line).
Note that the QSO shot-noise exceeds the cosmological power for all $k$.
For $z=1$ we plot the ELG shot-noise for an eBOSS-like sample,
assuming $b=1$.
At $z=1.5$ we plot the shot-noise assuming constant clustering and DESI-like
number densities.}
\label{fig:pk}
\end{figure}

\begin{figure*}
\begin{center}
\resizebox{\columnwidth}{!}{\includegraphics{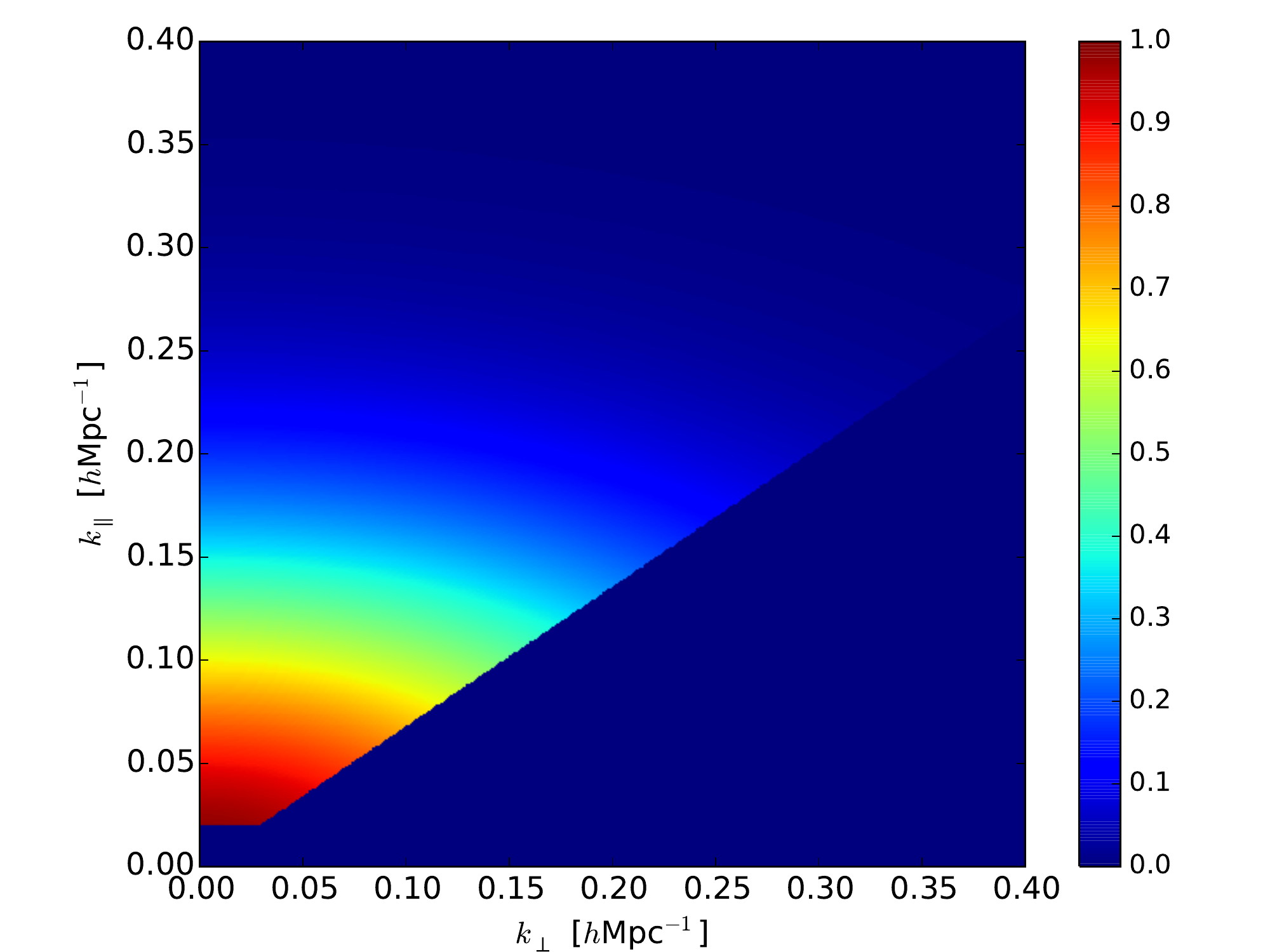}}
\resizebox{\columnwidth}{!}{\includegraphics{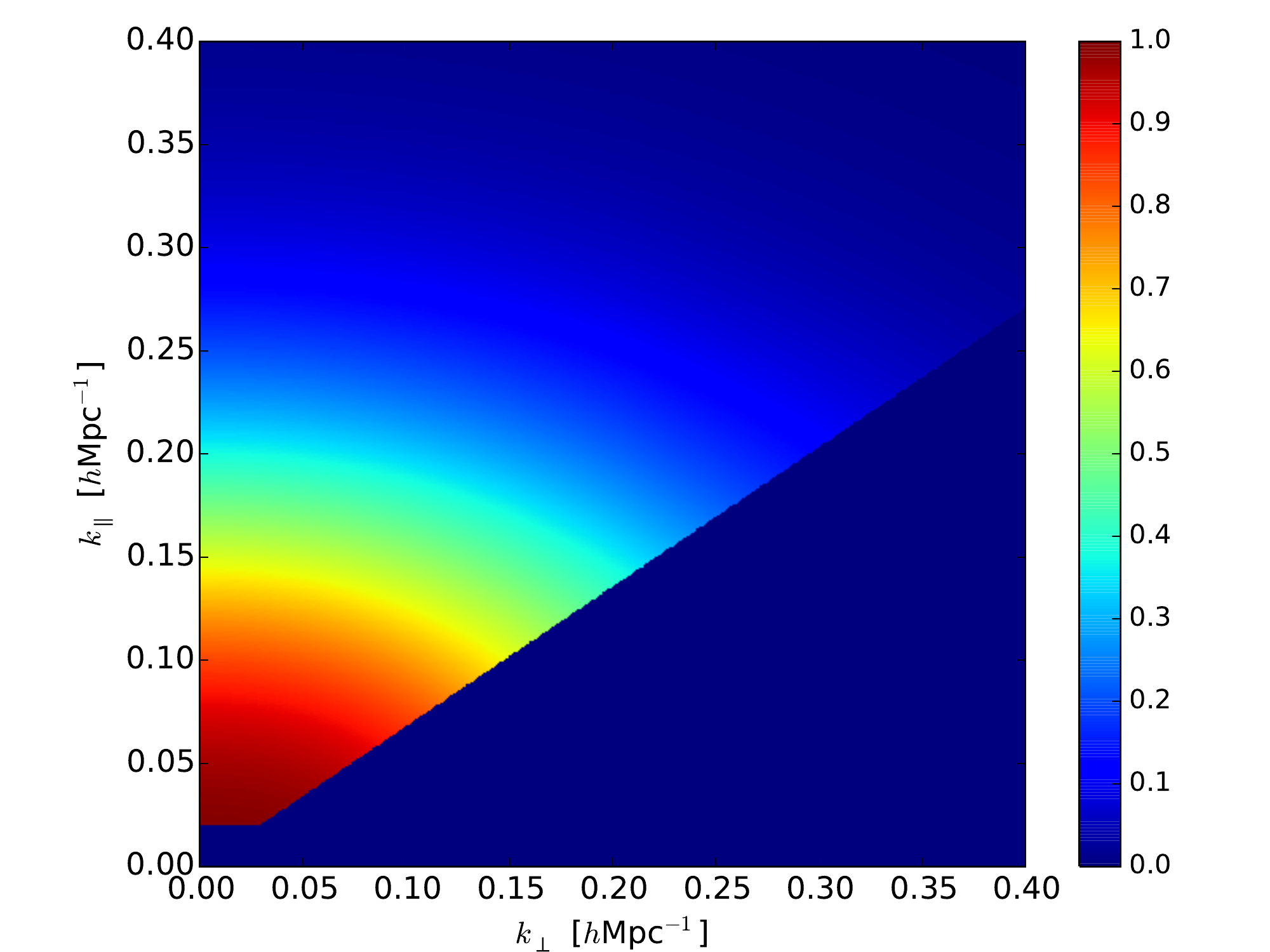}}
\resizebox{\columnwidth}{!}{\includegraphics{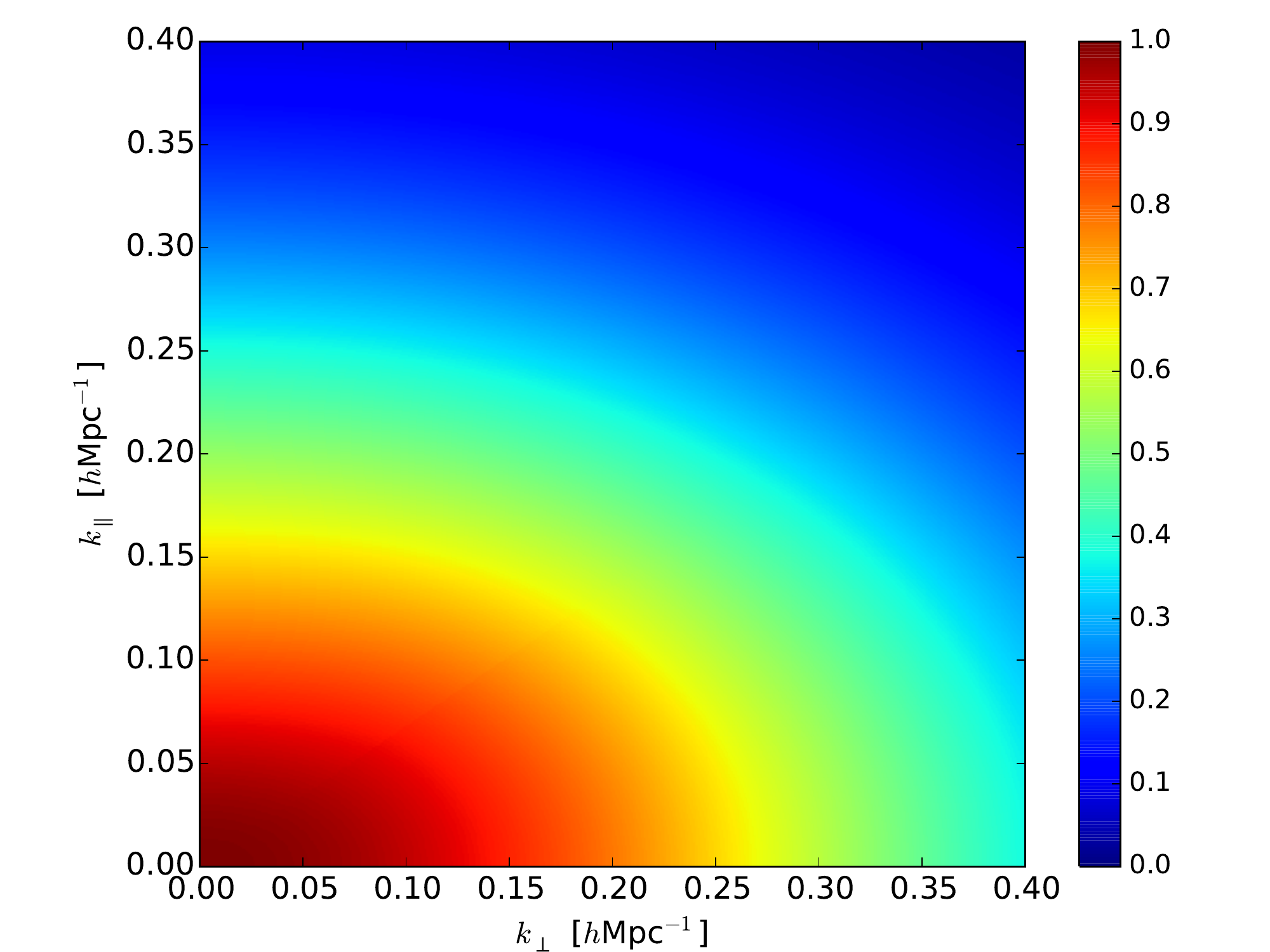}}
\resizebox{\columnwidth}{!}{\includegraphics{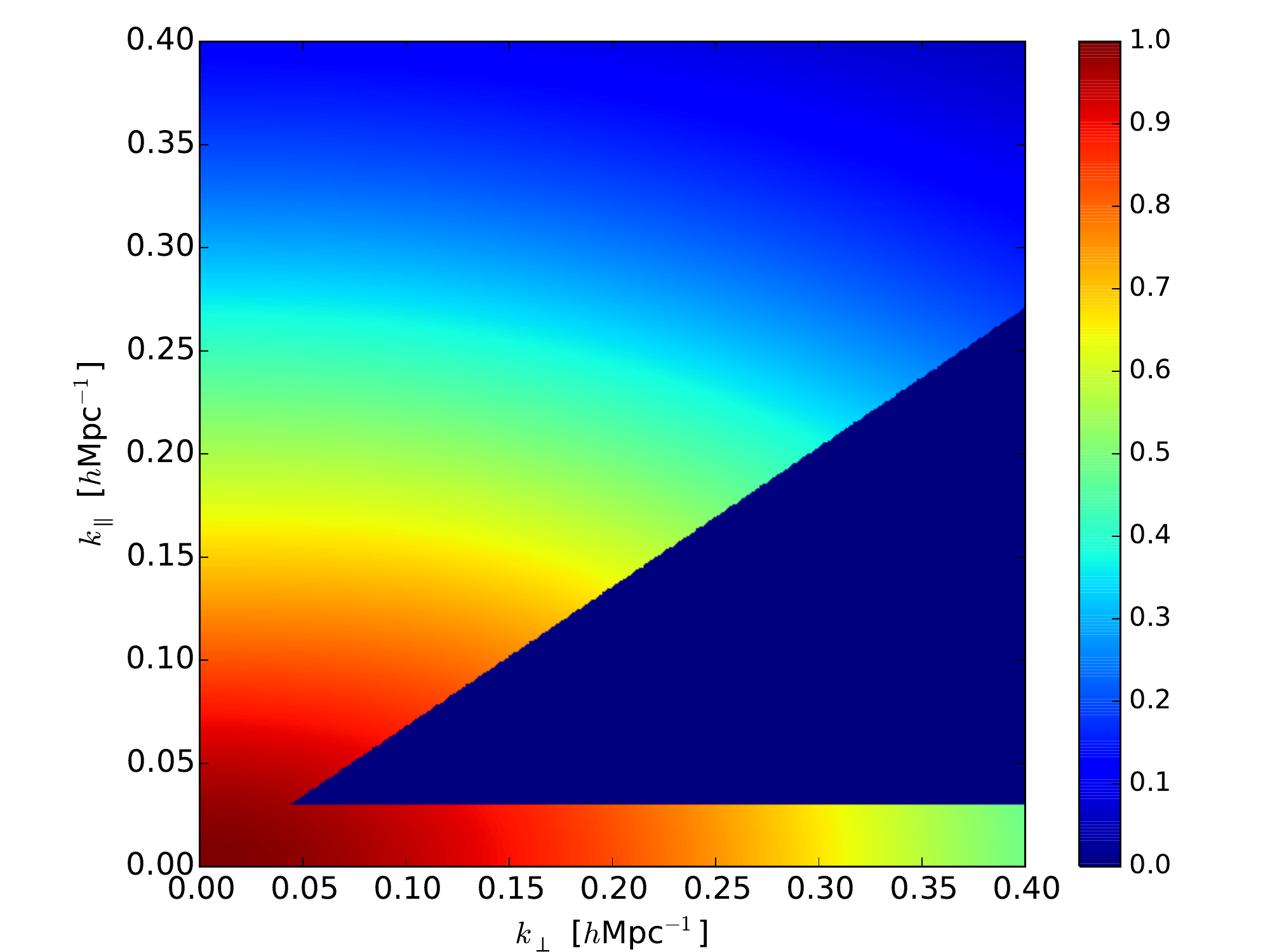}}
\end{center}
\caption{The damping function, $\mathcal{D}({\bf k})$, with $21\,$cm data
restricted to $k_\parallel\geq 0.02$ and $\mu\geq 0.56$ with an effective
density $b^2\bar{n} = 3\times 10^{-3}\,h^3{\rm Mpc}^{-3}$.
Top left: no reconstruction.
Top right: reconstruction only with $21\,$cm data.
Bottom left: reconstruction using ELGs with
$b^2\bar{n}=3\times 10^{-4}\,h^3\,{\rm Mpc}^{-3}$
to fill in the missing modes.
Bottom right: reconstruction using a photometric redshift survey with
$b^2\bar{n}=3\times 10^{-3}\,h^3\,{\rm Mpc}^{-3}$ to fill in the modes with
$k_\parallel<0.03\,h\,{\rm Mpc}^{-1}$.  To be able to access such modes
requires photo-$z$ precision of $\delta z/(1+z)\sim 0.01$ (see text).
Note that in spite of the wedge, non-missing modes along $k_\perp$ are much better
reconstructed because of redshift space distortions along $k_\parallel$.}
\label{fig:fullexample}
\end{figure*}

The top panel of Fig.~\ref{fig:missingmodes} demonstrates that loss of line-of-sight or
wedge modes at low $k$ significantly weakens the ability of a $21\,$cm
experiment to constrain the distance scale.
However, the modes which are missing are of very long wavelength, near
the peak of the CDM power spectrum, and thus can be measured relatively
well even by quite sparse tracers.
The combined density field then has a noise level which is set by the
$21\,$cm survey at high $k$ and the ``filler'' survey at lower $k$.

As a particular example we shall consider using ELGs\footnote{Surveys of
QSOs currently cover more area and have a wider redshift overlap with
planned $21\,$cm experiments, but are very sparse.  Assuming a fiducial
$b(z)=0.53+0.29(1+z)^2$ \citep{Cro05} and the number densities from
\citet{Daw15,Mye15} we find that $P_N\ge P_L$ for all $k$, see
Fig.~\ref{fig:pk}.  While this still allows high precision measurements of
$P(k)$, given enough volume, it leads to poor reconstruction.} as measured
for example by eBOSS or DESI.
Table 8 of \citet{Daw15} lists the ELG number density for the DECam ELG
sample as $\sim (1.5-3)\times 10^{-4}\,h^3{\rm Mpc}^{-3}$ over the
range $0.7<z<1$ and across 1,000 deg${}^2$.  We shall take the upper end
of this range as an optimistic example and assume the ELGs are unbiased
($b=1$).  This gives the ``filler'' data a number density an order of
magnitude lower than the effective number density for the $21\,$cm data.
The amplitudes of the shot noise and cosmological power, as a function
of $k$, are compared in Fig.~\ref{fig:pk}.

While in reality one would estimate the density field from the combination
of the surveys in an inverse variance manner, and the transition in the
noise is likely to be smooth with $\bk$, we shall instead take the noise
to be $1/\bar{n}_{21}$ where our $21\,$cm survey has data, sharply
transitioning to $1/\bar{n}_{ELG}$ where our $21\,$ survey does not.
[Our formalism can handle an arbitrary $P_N(\bk)$.]
The lower panel of Fig.~\ref{fig:missingmodes} shows the improvement
that such a survey combination would make in reconstruction -- in addition
to allowing a measurement of modes inside the foreground-dominated region
and so lowering the sample variance for those $k$s.

Fig.~\ref{fig:fullexample} shows the full $\mathcal{D}(k_\perp,k_\parallel)$
for four cases: no reconstruction, reconstruction using only $21\,$cm data,
reconstruction with the addition of the ELGs and reconstruction using a
photometric redshift sample (see below).
Even missing the modes indicated, a comparison of the top left and right
panels shows that reconstruction improves the signal-to-noise of the
acoustic signature.  However, the improvement that comes from including the
missing modes by using ELG data is dramatic, as a comparison of the bottom
left and top right panels shows.  We discuss the lower right panel below.

\begin{figure}
\begin{center}
\resizebox{\columnwidth}{!}{\includegraphics{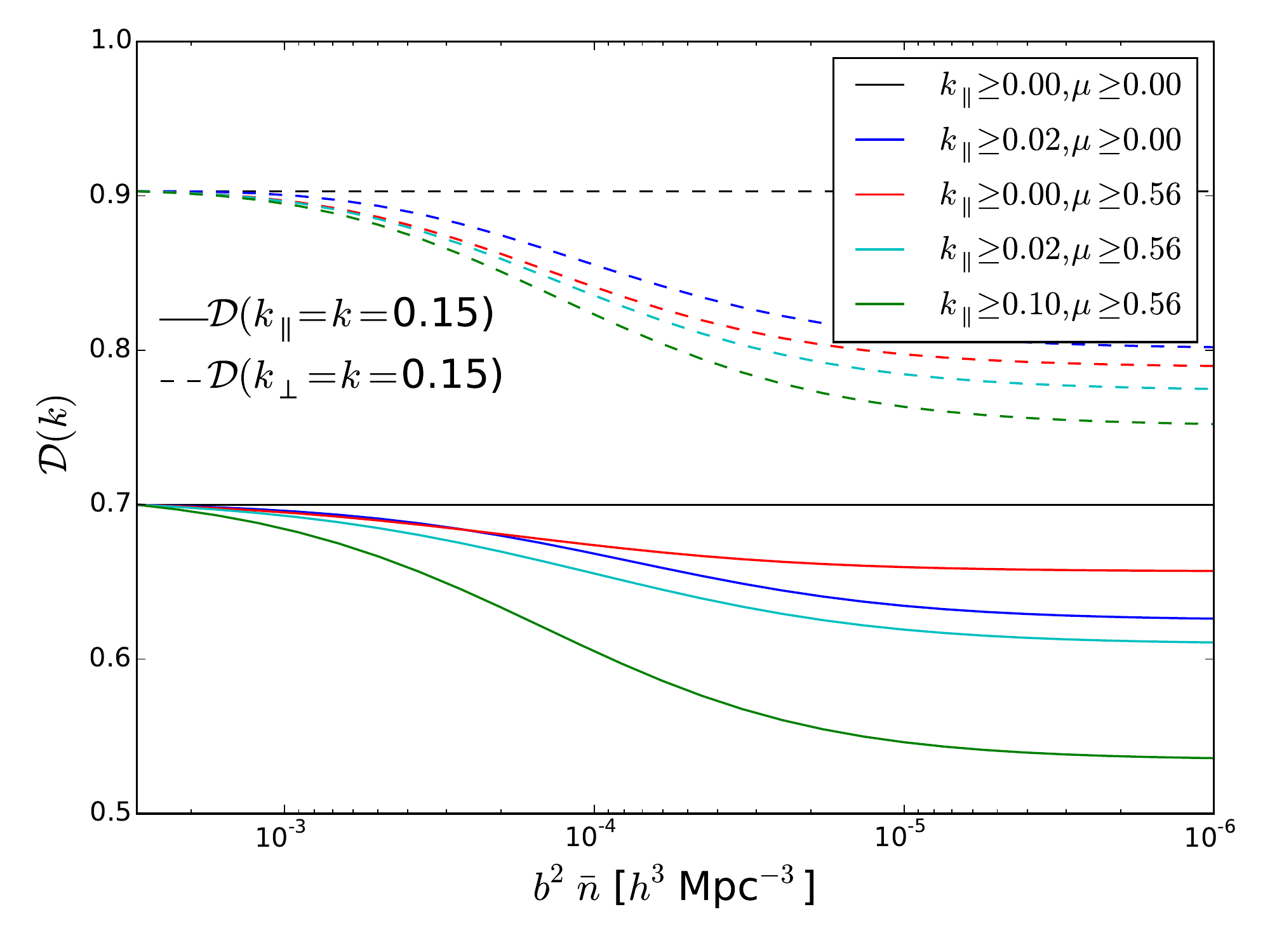}}
\end{center}
\caption{The damping function at $k=0.15\,h\,{\rm Mpc}^{-1}$ as a function
of ELG density for different wedge shapes (as indicated in the figure).
Solid lines show $\mathcal{D}$ for $k_\parallel = k = 0.15\,h\,{\rm Mpc}^{-1}$
while dashed lines show $k_\perp=k=0.15\,h\,{\rm Mpc}^{-1}$.
The small density, i.e., large noise, limits asymptotically reach the
$\mathcal{D}(k)$ where only the $21\,$cm measurements are available.
(If a mode is missing entirely when only $21\,$cm measurements are
available, then it will not contribute before or after
reconstruction.)
In spite of the wedge, modes along $k_\perp$ are much better reconstructed
because of redshift space distortions along $k_\parallel$.
The line for the constraint 
$k_\parallel \geq 0.1 \,h\, {\rm Mpc}^{-1}$ is the same for $\mu \geq 0,0.56$.
}
\label{fig:changenoise}
\end{figure}

We find that the combination of experiments does better than either does
alone.  The improvement over a $21\,$cm experiment which cannot measure
some modes is demonstrated above.  Although we do not show it here, the
combination of the $21\,$cm data and the ELGs provides slightly better
reconstruction than can be obtained with a survey of only ELGs
(unless their number density can be increased by an order of magnitude).
The damping scales, $\Sigma_{s}$ and $\Sigma_{ss}$, are common to all modes
and benefit from the regions of $k$-space with lower noise.
For our fiducial ELG number density the improvement is not dramatic since
the noise is subdominant to the signal power for a broad range of $k$,
however the situation changes if such high number densities cannot easily
be obtained over the desired redshift range.

To further explore the parameter space, Fig.~\ref{fig:changenoise} shows
the damping function along or transverse to the line of sight
(with $k=0.15\,h\,{\rm Mpc}^{-1}$) as a function of tracer noise
(specifically $b^2\bar{n}$) for different choices of the missing
modes.  One can see that once the number density of the tracer has
$b^2\bar{n}\simeq 10^{-4}\,h^3\,{\rm Mpc}^{-3}$ or larger it is able to
compensate for the modes missed by the $21\,$cm survey and improve the
performance of reconstruction. 
This characteristic number density is given by the cosmological power
($P_L$) at the scales where the $21\,$cm survey is missing modes, which
is close to the peak of the power spectrum.

So far our discussion has assumed that the ``filler'' sample comes from a
spectroscopic survey.  However, if only the low $k$ modes are needed, samples
and surveys with high-precision photometric redshifts could be used instead.
The lower right panel of Fig.~\ref{fig:fullexample} shows an optimistic case
where a photometric redshift survey can fill in all of the modes with
$k_\parallel<0.03\,h\,{\rm Mpc}^{-1}$.  (This includes some modes
already obtained from $21\,$cm outside the wedge, but adds information inside
the wedge.)
Translating a redshift uncertainty of $\delta z$ into a comoving distance
uncertainty of $\delta\chi=[c/H(z)]\delta z$, to probe
$k_\parallel=0.03\,h\,{\rm Mpc}^{-1}$ requires $\delta z/(1+z)<0.01$
at $z=1$.
Such photo-$z$ precision is in principle achievable, given enough filters.
As shown in Fig.~\ref{fig:fullexample} lower right, recovering
$k_\parallel<0.03\,h\,{\rm Mpc}^{-1}$ (for all $k_\perp$) can have a large
impact on reconstruction.  Conversely, if only
$k_\parallel<0.01\,h\,{\rm Mpc}^{-1}$ can be recovered then the gain
is minimal.

\section{Discussion}
\label{sec:discussion}

The study of large-scale structure has taught us a great deal about the
Universe in which we live and provides tight constraints on fundamental
physics.  One of the key observables in large-scale structure are the
so-called BAO, which provide a standard ruler enabling the measurement
of the expansion history of the Universe with low systematics and high
precision.  The BAO signal is degraded by non-linear evolution, but the
degradation can to some degree be overcome by density field reconstruction.

While $21\,$cm experiments can in principle measure large-scale structure
efficiently at very high redshifts, where galaxy redshift surveys become
increasingly expensive, they may suffer from large foreground contamination
in a ``wedge'' in $k$-space.  The modes lost due to foregrounds have an
impact on BAO measurements but an even larger impact upon reconstruction
\citep[as recently emphasized by][]{SeoHir15}.
In this paper we have demonstrated, using an analytic model based on
low-order Lagrangian perturbation theory, that even a relatively sparse
tracer of the density field can ``fill in'' the missing modes.
The combined data can be more powerful than either of its parts: reaping the
benefits of full $k$-coverage and low shot-noise.  This allows density field
reconstruction and tight measurements at smaller scales, which can be
especially useful at higher $z$.

We have presented formulae for the efficacy of reconstruction in the case
of biased tracers of the density field, including redshift-space distortions,
anisotropic noise and noise-filtering during reconstruction.
The missing modes can be modeled as infinite noise, while the effects
of having two different samples can be modeled via anisotropic noise.
These extend the formulae in the literature, but agree with those formulae
in the appropriate limits.  We find that the details of how BAO are predicted
to be damped in Lagrangian perturbation theory depend upon the shape of the
filter applied to estimate the shift field.  If this finding holds up in
simulations, it opens the possibility of ``shaping'' the filter to improve
the performance of reconstruction.

While even a very noisy tracer of the density field can be used to
measure the power spectrum if enough modes can be averaged together,
our intuition tells us that reconstruction --- which depends on the 3D
density field and not just its power spectrum --- requires noise per
$k$-mode less than the cosmological signal.
Our calculation quantifies and supports this intuition, and we show that
to measure the low-$k$ modes missed by a $21\,$cm survey tracers with
$b^2\bar{n}\sim 10^{-3}-10^{-4}\,h^3\,{\rm Mpc}^{-3}$ are required.

Of the existing surveys that cover large cosmological volumes at high
redshift, the QSO surveys are currently too sparse to be of significant
benefit.  An increase in number density, however, would lead to improved
performance.
Some of the low $k_\parallel$ modes could, in principle, be filled in with
a photometric sample with excellent photometric redshifts (obtained, perhaps,
using multiple medium bands).
In the near term the most promising tracer, in the redshift range $z>1$ where
most future $21\,$cm surveys will be operating, are ELGs such as will be
measured by eBOSS or DESI.
If eBOSS can achieve its forecast number densities it would significantly
improve reconstruction for $21\,$cm surveys which overlap in volume.
In the future even the higher $z$ tail of the DESI ELG sample (with its
higher bias) would be beneficial to $21\,$cm surveys which are unable to
work deep into the foreground wedge.

\vspace{0.2in}
We would like to thank Marcel Schmittfull for helpful conversations on
acoustic oscillations and reconstruction, and Josh Dillon, Daniel
Eisenstein, Marcel Schmittfull, Uros Seljak and 
Hee-Jong Seo for helpful feedback on the draft, and the anonymous
referee for additional helpful suggestions.
This work was begun at the Aspen Center for Physics, which is supported
by National Science Foundation grant PHY-1066293.  We thank the Center
for its hospitality. T.-C.C. acknowledges support from MoST grant
103-2112-M-001-002-MY3. 
This work made extensive use of the NASA Astrophysics Data System and
of the {\tt astro-ph} preprint archive at {\tt arXiv.org}.


\newcommand{\aj}{AJ}
\newcommand{\apj}{ApJ}
\newcommand{\apjs}{ApJ Suppl.}
\newcommand{\mnras}{MNRAS}
\newcommand{\araa}{ARA{\&}A}
\newcommand{\aap}{A{\&}A}
\newcommand{\pre}{PRE}
\newcommand{\prd}{Phys. Rev. D}
\newcommand{\apjl}{ApJL}
\newcommand{\physrep}{Physics Reports}
\newcommand{\nat}{Nature}

\bibliographystyle{mn2e}
\bibliography{ms}

\begin{thebibliography}{54}
\expandafter\ifx\csname natexlab\endcsname\relax\def\natexlab#1{#1}\fi

\bibitem[{{Ali} \& {Bharadwaj}(2014)}]{AliBha14}
{Ali} S.~S., {Bharadwaj} S., 2014, Journal of Astrophysics and Astronomy, 35,
  157

\bibitem[{{Anderson} {et~al}\mbox{.}(2014){Anderson}, {Aubourg}, {Bailey},
  {Beutler}, {Bhardwaj}, {Blanton}, {Bolton}, {Brinkmann}, {Brownstein},
  {Burden}, {Chuang}, {Cuesta}, {Dawson}, {Eisenstein}, {Escoffier}, {Gunn},
  {Guo}, {Ho}, {Honscheid}, {Howlett}, {Kirkby}, {Lupton}, {Manera},
  {Maraston}, {McBride}, {Mena}, {Montesano}, {Nichol}, {Nuza}, {Olmstead},
  {Padmanabhan}, {Palanque-Delabrouille}, {Parejko}, {Percival}, {Petitjean},
  {Prada}, {Price-Whelan}, {Reid}, {Roe}, {Ross}, {Ross}, {Sabiu}, {Saito},
  {Samushia}, {S{\'a}nchez}, {Schlegel}, {Schneider}, {Scoccola}, {Seo},
  {Skibba}, {Strauss}, {Swanson}, {Thomas}, {Tinker}, {Tojeiro}, {Maga{\~n}a},
  {Verde}, {Wake}, {Weaver}, {Weinberg}, {White}, {Xu}, {Y{\`e}che}, {Zehavi},
  \& {Zhao}}]{And14}
{Anderson} L. {et~al.}, 2014, \mnras, 441, 24

\bibitem[{{Ansari} {et~al}\mbox{.}(2012){Ansari}, {Campagne}, {Colom},
  {Magneville}, {Martin}, {Moniez}, {Rich}, \& {Y{\`e}che}}]{Ans12}
{Ansari} R., {Campagne} J.-E., {Colom} P., {Magneville} C., {Martin} J.-M.,
  {Moniez} M., {Rich} J., {Y{\`e}che} C., 2012, Comptes Rendus Physique, 13, 46

\bibitem[{{Bharadwaj}(1996)}]{Bha96}
{Bharadwaj} S., 1996, \apj, 460, 28

\bibitem[{{Burden} {et~al}\mbox{.}(2014){Burden}, {Percival}, {Manera},
  {Cuesta}, {Vargas Magana}, \& {Ho}}]{Bur14}
{Burden} A., {Percival} W.~J., {Manera} M., {Cuesta} A.~J., {Vargas Magana} M.,
  {Ho} S., 2014, \mnras, 445, 3152

\bibitem[{{Busca} {et~al}\mbox{.}(2013){Busca}, {Delubac}, {Rich}, {Bailey},
  {Font-Ribera}, {Kirkby}, {Le Goff}, {Pieri}, {Slosar}, {Aubourg}, {Bautista},
  {Bizyaev}, {Blomqvist}, {Bolton}, {Bovy}, {Brewington}, {Borde}, {Brinkmann},
  {Carithers}, {Croft}, {Dawson}, {Ebelke}, {Eisenstein}, {Hamilton}, {Ho},
  {Hogg}, {Honscheid}, {Lee}, {Lundgren}, {Malanushenko}, {Malanushenko},
  {Margala}, {Maraston}, {Mehta}, {Miralda-Escud{\'e}}, {Myers}, {Nichol},
  {Noterdaeme}, {Olmstead}, {Oravetz}, {Palanque-Delabrouille}, {Pan},
  {P{\^a}ris}, {Percival}, {Petitjean}, {Roe}, {Rollinde}, {Ross}, {Rossi},
  {Schlegel}, {Schneider}, {Shelden}, {Sheldon}, {Simmons}, {Snedden},
  {Tinker}, {Viel}, {Weaver}, {Weinberg}, {White}, {Y{\`e}che}, \&
  {York}}]{Bus13}
{Busca} N.~G. {et~al.}, 2013, \aap, 552, A96

\bibitem[{{Carlson}, {Reid} \& {White}(2013){Carlson}, {Reid}, \&
  {White}}]{CarReiWhi13}
{Carlson} J., {Reid} B., {White} M., 2013, \mnras, 429, 1674

\bibitem[{{Chang} {et~al}\mbox{.}(2010){Chang}, {Pen}, {Bandura}, \&
  {Peterson}}]{Cha10}
{Chang} T.-C., {Pen} U.-L., {Bandura} K., {Peterson} J.~B., 2010, Nature, 466,
  463

\bibitem[{{Chang} {et~al}\mbox{.}(2008){Chang}, {Pen}, {Peterson}, \&
  {McDonald}}]{Cha08}
{Chang} T.-C., {Pen} U.-L., {Peterson} J.~B., {McDonald} P., 2008, Physical
  Review Letters, 100, 091303

\bibitem[{{Crocce} \& {Scoccimarro}(2008)}]{CroSco08}
{Crocce} M., {Scoccimarro} R., 2008, \prd, 77, 023533

\bibitem[{{Croom} {et~al}\mbox{.}(2005){Croom}, {Boyle}, {Shanks}, {Smith},
  {Miller}, {Outram}, {Loaring}, {Hoyle}, \& {da {\^A}ngela}}]{Cro05}
{Croom} S.~M. {et~al.}, 2005, \mnras, 356, 415

\bibitem[{{Datta}, {Bowman} \& {Carilli}(2010){Datta}, {Bowman}, \&
  {Carilli}}]{DatBowCar10}
{Datta} A., {Bowman} J.~D., {Carilli} C.~L., 2010, \apj, 724, 526

\bibitem[{{Dawson} {et~al}\mbox{.}(2015){Dawson}, {Kneib}, {Percival}, {Alam},
  {Albareti}, {Anderson}, {Armengaud}, {Aubourg}, {Bailey}, {Bautista},
  {Berlind}, {Bershady}, {Beutler}, {Bizyaev}, {Blanton}, {Blomqvist},
  {Bolton}, {Bovy}, {Brandt}, {Brinkmann}, {Brownstein}, {Burtin}, {Busca},
  {Cai}, {Chuang}, {Clerc}, {Comparat}, {Cope}, {Croft}, {Cruz-Gonzalez}, {da
  Costa}, {Cousinou}, {Darling}, {de la Torre}, {Delubac}, {du Mas des
  Bourboux}, {Dwelly}, {Ealet}, {Eisenstein}, {Eracleous}, {Escoffier}, {Fan},
  {Finoguenov}, {Font-Ribera}, {Frinchaboy}, {Gaulme}, {Georgakakis}, {Green},
  {Guo}, {Guy}, {Ho}, {Holder}, {Huehnerhoff}, {Hutchinson}, {Jing}, {Jullo},
  {Kamble}, {Kinemuchi}, {Kirkby}, {Kitaura}, {Klaene}, {Laher}, {Lang},
  {Laurent}, {Le Goff}, {Li}, {Liang}, {Lima}, {Lin}, {Lin}, {Lin}, {Long},
  {Lundgren}, {MacDonald}, {Geimba Maia}, {Malanushenko}, {Malanushenko},
  {Mariappan}, {McBride}, {McGreer}, {Menard}, {Merloni}, {Meza},
  {Montero-Dorta}, {Muna}, {Myers}, {Nandra}, {Naugle}, {Newman}, {Noterdaeme},
  {Nugent}, {Ogando}, {Olmstead}, {Oravetz}, {Oravetz}, {Padmanabhan},
  {Palanque-Delabrouille}, {Pan}, {Parejko}, {Paris}, {Peacock}, {Petitjean},
  {Pieri}, {Pisani}, {Prada}, {Prakash}, {Raichoor}, {Reid}, {Rich}, {Ridl},
  {Rodriguez-Torres}, {Carnero Rosell}, {Ross}, {Rossi}, {Ruan}, {Salvato},
  {Sayres}, {Schneider}, {Schlegel}, {Seljak}, {Seo}, {Sesar}, {Shandera},
  {Shu}, {Slosar}, {Sobreira}, {Strauss}, {Streblyanska}, {Suzuki}, {Tao},
  {Tinker}, {Tojeiro}, {Vargas-Magana}, {Wang}, {Weaver}, {Weinberg}, {White},
  {Wood-Vasey}, {Yeche}, {Zhai}, {Zhao}, {Zhao}, {Zheng}, {Ben Zhu}, \&
  {Zou}}]{Daw15}
{Dawson} K.~S. {et~al.}, 2015, ArXiv e-prints

\bibitem[{{Delubac} {et~al}\mbox{.}(2015){Delubac}, {Bautista}, {Busca},
  {Rich}, {Kirkby}, {Bailey}, {Font-Ribera}, {Slosar}, {Lee}, {Pieri},
  {Hamilton}, {Aubourg}, {Blomqvist}, {Bovy}, {Brinkmann}, {Carithers},
  {Dawson}, {Eisenstein}, {Gontcho}, {Kneib}, {Le Goff}, {Margala},
  {Miralda-Escud{\'e}}, {Myers}, {Nichol}, {Noterdaeme}, {O'Connell},
  {Olmstead}, {Palanque-Delabrouille}, {P{\^a}ris}, {Petitjean}, {Ross},
  {Rossi}, {Schlegel}, {Schneider}, {Weinberg}, {Y{\`e}che}, \& {York}}]{Del15}
{Delubac} T. {et~al.}, 2015, \aap, 574, A59

\bibitem[{{Eisenstein} {et~al}\mbox{.}(2007){Eisenstein}, {Seo}, {Sirko}, \&
  {Spergel}}]{Eis07}
{Eisenstein} D.~J., {Seo} H.-J., {Sirko} E., {Spergel} D.~N., 2007, \apj, 664,
  675

\bibitem[{{Eisenstein}, {Seo} \& {White}(2007){Eisenstein}, {Seo}, \&
  {White}}]{ESW07}
{Eisenstein} D.~J., {Seo} H.-J., {White} M., 2007, \apj, 664, 660

\bibitem[{{Font-Ribera} {et~al}\mbox{.}(2014){Font-Ribera}, {Kirkby}, {Busca},
  {Miralda-Escud{\'e}}, {Ross}, {Slosar}, {Rich}, {Aubourg}, {Bailey},
  {Bhardwaj}, {Bautista}, {Beutler}, {Bizyaev}, {Blomqvist}, {Brewington},
  {Brinkmann}, {Brownstein}, {Carithers}, {Dawson}, {Delubac}, {Ebelke},
  {Eisenstein}, {Ge}, {Kinemuchi}, {Lee}, {Malanushenko}, {Malanushenko},
  {Marchante}, {Margala}, {Muna}, {Myers}, {Noterdaeme}, {Oravetz},
  {Palanque-Delabrouille}, {P{\^a}ris}, {Petitjean}, {Pieri}, {Rossi},
  {Schneider}, {Simmons}, {Viel}, {Yeche}, \& {York}}]{Fon14}
{Font-Ribera} A. {et~al.}, 2014, JCAP, 5, 27

\bibitem[{{Furlanetto}, {Oh} \& {Briggs}(2006){Furlanetto}, {Oh}, \&
  {Briggs}}]{Fur06}
{Furlanetto} S.~R., {Oh} S.~P., {Briggs} F.~H., 2006, \physrep, 433, 181

\bibitem[{{Kirkby} {et~al}\mbox{.}(2013){Kirkby}, {Margala}, {Slosar},
  {Bailey}, {Busca}, {Delubac}, {Rich}, {Bautista}, {Blomqvist}, {Brownstein},
  {Carithers}, {Croft}, {Dawson}, {Font-Ribera}, {Miralda-Escud{\'e}}, {Myers},
  {Nichol}, {Palanque-Delabrouille}, {P{\^a}ris}, {Petitjean}, {Rossi},
  {Schlegel}, {Schneider}, {Viel}, {Weinberg}, \& {Y{\`e}che}}]{Kir13}
{Kirkby} D. {et~al.}, 2013, JCAP, 3, 24

\bibitem[{{Liu}, {Parsons} \& {Trott}(2014){Liu}, {Parsons}, \&
  {Trott}}]{LiuParTro14}
{Liu} A., {Parsons} A.~R., {Trott} C.~M., 2014, \prd, 90, 023019

\bibitem[{{Matsubara}(2008)}]{Mat08}
{Matsubara} T., 2008, \prd, 77, 063530

\bibitem[{{McCullagh} \& {Szalay}(2012)}]{McCSza12}
{McCullagh} N., {Szalay} A.~S., 2012, \apj, 752, 21

\bibitem[{{Meiksin}, {White} \& {Peacock}(1999){Meiksin}, {White}, \&
  {Peacock}}]{MeiWhiPea99}
{Meiksin} A., {White} M., {Peacock} J.~A., 1999, \mnras, 304, 851

\bibitem[{{Morales} {et~al}\mbox{.}(2012){Morales}, {Hazelton}, {Sullivan}, \&
  {Beardsley}}]{Mor12}
{Morales} M.~F., {Hazelton} B., {Sullivan} I., {Beardsley} A., 2012, \apj, 752,
  137

\bibitem[{{Myers} {et~al}\mbox{.}(2015){Myers}, {Palanque-Delabrouille},
  {Prakash}, {P{\^a}ris}, {Yeche}, {Dawson}, {Bovy}, {Lang}, {Schlegel},
  {Newman}, {Petitjean}, {Kneib}, {Laurent}, {Percival}, {Ross}, {Seo},
  {Tinker}, {Armengaud}, {Brownstein}, {Burtin}, {Cai}, {Comparat}, {Kasliwal},
  {Kulkarni}, {Laher}, {Levitan}, {McBride}, {McGreer}, {Miller}, {Nugent},
  {Ofek}, {Rossi}, {Ruan}, {Schneider}, {Sesar}, {Streblyanska}, {Surace}, \&
  {for the SDSS-IV/eBOSS collaboration}}]{Mye15}
{Myers} A.~D. {et~al.}, 2015, ArXiv e-prints

\bibitem[{{Nan} {et~al}\mbox{.}(2011){Nan}, {Li}, {Jin}, {Wang}, {Zhu}, {Zhu},
  {Zhang}, {Yue}, \& {Qian}}]{Nan11}
{Nan} R. {et~al.}, 2011, International Journal of Modern Physics D, 20, 989

\bibitem[{{Noh}, {White} \& {Padmanabhan}(2009){Noh}, {White}, \&
  {Padmanabhan}}]{NohWhiPad09}
{Noh} Y., {White} M., {Padmanabhan} N., 2009, \prd, 80, 123501

\bibitem[{Olive {et~al}\mbox{.}(2014)Olive {et~al.}}]{PDG}
Olive K.~A., {et~al.}, 2014, Chin. Phys., C38, 090001

\bibitem[{{Padmanabhan} \& {White}(2009)}]{PadWhi09}
{Padmanabhan} N., {White} M., 2009, \prd, 80, 063508

\bibitem[{{Padmanabhan}, {White} \& {Cohn}(2009){Padmanabhan}, {White}, \&
  {Cohn}}]{PadWhiCoh09}
{Padmanabhan} N., {White} M., {Cohn} J.~D., 2009, \prd, 79, 063523

\bibitem[{{Padmanabhan} {et~al}\mbox{.}(2012){Padmanabhan}, {Xu}, {Eisenstein},
  {Scalzo}, {Cuesta}, {Mehta}, \& {Kazin}}]{Pad12}
{Padmanabhan} N., {Xu} X., {Eisenstein} D.~J., {Scalzo} R., {Cuesta} A.~J.,
  {Mehta} K.~T., {Kazin} E., 2012, \mnras, 427, 2132

\bibitem[{{Parsons} {et~al}\mbox{.}(2012){Parsons}, {Pober}, {Aguirre},
  {Carilli}, {Jacobs}, \& {Moore}}]{Par12}
{Parsons} A.~R., {Pober} J.~C., {Aguirre} J.~E., {Carilli} C.~L., {Jacobs}
  D.~C., {Moore} D.~F., 2012, \apj, 756, 165

\bibitem[{{Pober}(2015)}]{Pob15}
{Pober} J.~C., 2015, \mnras, 447, 1705

\bibitem[{{Pober} {et~al}\mbox{.}(2013){Pober}, {Parsons}, {DeBoer},
  {McDonald}, {McQuinn}, {Aguirre}, {Ali}, {Bradley}, {Chang}, \&
  {Morales}}]{Pob13}
{Pober} J.~C. {et~al.}, 2013, \aj, 145, 65

\bibitem[{{Schmittfull} {et~al}\mbox{.}(2015){Schmittfull}, {Feng}, {Beutler},
  {Sherwin}, \& {Yat Chu}}]{Sch15}
{Schmittfull} M., {Feng} Y., {Beutler} F., {Sherwin} B., {Yat Chu} M., 2015,
  ArXiv e-prints

\bibitem[{{Seo} {et~al}\mbox{.}(2015){Seo}, {Beutler}, {Ross}, \&
  {Saito}}]{Seo15}
{Seo} H.-J., {Beutler} F., {Ross} A.~J., {Saito} S., 2015, ArXiv e-prints

\bibitem[{{Seo} {et~al}\mbox{.}(2010{\natexlab{a}}){Seo}, {Dodelson},
  {Marriner}, {Mcginnis}, {Stebbins}, {Stoughton}, \& {Vallinotto}}]{Seo10b}
{Seo} H.-J., {Dodelson} S., {Marriner} J., {Mcginnis} D., {Stebbins} A.,
  {Stoughton} C., {Vallinotto} A., 2010{\natexlab{a}}, \apj, 721, 164

\bibitem[{{Seo} {et~al}\mbox{.}(2010{\natexlab{b}}){Seo}, {Eckel},
  {Eisenstein}, {Mehta}, {Metchnik}, {Padmanabhan}, {Pinto}, {Takahashi},
  {White}, \& {Xu}}]{Seo10}
{Seo} H.-J. {et~al.}, 2010{\natexlab{b}}, \apj, 720, 1650

\bibitem[{{Seo} \& {Hirata}(2015)}]{SeoHir15}
{Seo} H.-J., {Hirata} C.~M., 2015, ArXiv e-prints

\bibitem[{{Shaw} {et~al}\mbox{.}(2014){Shaw}, {Sigurdson}, {Pen}, {Stebbins},
  \& {Sitwell}}]{Sha14}
{Shaw} J.~R., {Sigurdson} K., {Pen} U.-L., {Stebbins} A., {Sitwell} M., 2014,
  \apj, 781, 57

\bibitem[{{Shaw} {et~al}\mbox{.}(2015){Shaw}, {Sigurdson}, {Sitwell},
  {Stebbins}, \& {Pen}}]{Sha15}
{Shaw} J.~R., {Sigurdson} K., {Sitwell} M., {Stebbins} A., {Pen} U.-L., 2015,
  \prd, 91, 083514

\bibitem[{{Sherwin} \& {Zaldarriaga}(2012)}]{SheZal12}
{Sherwin} B.~D., {Zaldarriaga} M., 2012, \prd, 85, 103523

\bibitem[{{Slosar} {et~al}\mbox{.}(2013){Slosar}, {Ir{\v s}i{\v c}}, {Kirkby},
  {Bailey}, {Busca}, {Delubac}, {Rich}, {Aubourg}, {Bautista}, {Bhardwaj},
  {Blomqvist}, {Bolton}, {Bovy}, {Brownstein}, {Carithers}, {Croft}, {Dawson},
  {Font-Ribera}, {Le Goff}, {Ho}, {Honscheid}, {Lee}, {Margala}, {McDonald},
  {Medolin}, {Miralda-Escud{\'e}}, {Myers}, {Nichol}, {Noterdaeme},
  {Palanque-Delabrouille}, {P{\^a}ris}, {Petitjean}, {Pieri}, {Pi{\v s}kur},
  {Roe}, {Ross}, {Rossi}, {Schlegel}, {Schneider}, {Suzuki}, {Sheldon},
  {Seljak}, {Viel}, {Weinberg}, \& {Y{\`e}che}}]{Slo13}
{Slosar} A. {et~al.}, 2013, JCAP, 4, 26

\bibitem[{{Tassev} \& {Zaldarriaga}(2012{\natexlab{a}})}]{TasZal12a}
{Tassev} S., {Zaldarriaga} M., 2012{\natexlab{a}}, JCAP, 4, 13

\bibitem[{{Tassev} \& {Zaldarriaga}(2012{\natexlab{b}})}]{TasZal12b}
{Tassev} S., {Zaldarriaga} M., 2012{\natexlab{b}}, JCAP, 10, 6

\bibitem[{{Taylor} \& {Hamilton}(1996)}]{TayHam96}
{Taylor} A.~N., {Hamilton} A.~J.~S., 1996, \mnras, 282, 767

\bibitem[{{Tojeiro} {et~al}\mbox{.}(2014){Tojeiro}, {Ross}, {Burden},
  {Samushia}, {Manera}, {Percival}, {Beutler}, {Brinkmann}, {Brownstein},
  {Cuesta}, {Dawson}, {Eisenstein}, {Ho}, {Howlett}, {McBride}, {Montesano},
  {Olmstead}, {Parejko}, {Reid}, {S{\'a}nchez}, {Schlegel}, {Schneider},
  {Tinker}, {Maga{\~n}a}, \& {White}}]{Toj14}
{Tojeiro} R. {et~al.}, 2014, \mnras, 440, 2222

\bibitem[{{Vargas-Maga{\~n}a} {et~al}\mbox{.}(2014){Vargas-Maga{\~n}a}, {Ho},
  {Xu}, {S{\'a}nchez}, {O'Connell}, {Eisenstein}, {Cuesta}, {Percival}, {Ross},
  {Aubourg}, {Brownstein}, {Escoffier}, {Kirkby}, {Manera}, {Schneider},
  {Tinker}, \& {Weaver}}]{Var14}
{Vargas-Maga{\~n}a} M. {et~al.}, 2014, \mnras, 445, 2

\bibitem[{{Vedantham}, {Udaya Shankar} \& {Subrahmanyan}(2012){Vedantham},
  {Udaya Shankar}, \& {Subrahmanyan}}]{Ved12}
{Vedantham} H., {Udaya Shankar} N., {Subrahmanyan} R., 2012, \apj, 745, 176

\bibitem[{{White}(2010)}]{Whi10}
{White} M., 2010, ArXiv e-prints

\bibitem[{{White}(2015)}]{Whi15}
{White} M., 2015, \mnras, 450, 3822

\bibitem[{{Xu} {et~al}\mbox{.}(2013){Xu}, {Cuesta}, {Padmanabhan},
  {Eisenstein}, \& {McBride}}]{Xu13}
{Xu} X., {Cuesta} A.~J., {Padmanabhan} N., {Eisenstein} D.~J., {McBride} C.~K.,
  2013, \mnras, 431, 2834

\bibitem[{{Zel'dovich}(1970)}]{Zel70}
{Zel'dovich} Y.~B., 1970, \aap, 5, 84

\bibitem[{{Zhu} {et~al}\mbox{.}(2015){Zhu}, {Pen}, {Yu}, {Er}, \&
  {Chen}}]{Zhu15}
{Zhu} H.-M., {Pen} U.-L., {Yu} Y., {Er} X., {Chen} X., 2015, ArXiv e-prints

\end{thebibliography}

\begin{appendix}

\section{Deriving the damping}
\label{app:derivation}

Starting with the expression for $\delta(\bk)$ in the text
[Eq.~\ref{eqn:deltak}] we can write [setting bias $b=1$, or
$F[\lambda]\propto\delta^{(D)}(\lambda)$]
\begin{equation}
  \left\langle\delta(\bk_1)\delta^\star(\bk_2)\right\rangle =
  \left\langle\int d\bq_1\,d\bq_2   e^{-i\bk_1\cdot\bq_1} e^{i\bk_2\cdot\bq_2}
  [e^{-i\bk_1\cdot \bP(\bq_1)} e^{i\bk_2\cdot\bP(\bq_2)} -1]
  \right\rangle \quad .
\end{equation}
Employing the cumulant theorem for a Gaussian variable
($\langle\exp[iX]\rangle=\exp[-(1/2)\langle X^2\rangle]$),
making a change of variables $d\bq_1\,d\bq_2\to d\mathbf{Q}\,d\bq$ with
$\bq=\bq_1-\bq_2$ and $2\mathbf{Q}=\bq_1+\bq_2$ and using translational
symmetry we have
\begin{equation}
  P(k) = \int d^3q\ e^{-i\bk\cdot\bq}\exp\left[
  -k_ik_j\left\{\xi_{ij}(\mathbf{0})-\xi_{ij}(\bq)\right\}\right]
\end{equation}
where we have defined
$\xi_{ij}(\bq)=\left\langle\Psi_i(\bq)\Psi_j(\mathbf{0})\right\rangle$
and assumed $\bk\ne 0$.
Extracting the $q$-independent piece of the integral and expanding the
exponential gives
\begin{equation}
  P(k) = \exp\left[-k_ik_j\xi_{ij}(\mathbf{0})\right]
  \int d^3q\ e^{-i\bk\cdot\bq}\left[k_ik_j\xi_{ij}(\bq)+\cdots\right]
\end{equation}
The Fourier transform of the $k_ik_j\xi_{ij}(\bq)$ term can easily be
shown to be $P_L(k)$ to lowest order,
\begin{eqnarray}
  \int d^3q\ e^{-i\bk\cdot\bq} k_ik_j\xi_{ij}(\bq) &=&
  k_ik_j\int d^3q\ e^{-i\bk\cdot\bq}\int\frac{d^3p}{(2\pi)^3} \nonumber \\
  &\times& e^{i\bp\cdot\bq}\frac{p_ip_j}{p^4}P_L(p) \\
  &=& k_ik_j\int\frac{d^3p}{(2\pi)^3}\frac{p_ip_j}{p^4}P_L(p) \nonumber \\
  &\times& \ (2\pi)^3\delta(\bp-\bk) \\
  &=& P_L(k)
\end{eqnarray}
while the exponential is the source of the damping term of the BAO
features (or the broadening of the acoustic peak in the correlation
function)
\begin{equation}
  \xi_{ij}(\mathbf{0}) = \int\frac{d^3k}{(2\pi)^3}\frac{k_ik_j}{k^4}P_L(k)
  = \frac{\delta_{ij}}{3}\int \frac{dk}{2\pi^2}\ P_L(k)
  \quad .
\end{equation}

\section{Reconstruction in Eulerian perturbation theory}
\label{app:eulerian}

Most of the reconstruction literature either prescribes its implementation
on a data set or interprets this implementation within Lagrangian
perturbation theory.
However, recently \citet{Sch15} developed a theory of reconstruction
based on Eulerian perturbation theory and introduced several new
reconstruction schemes.
One advantage of the Eulerian formulation,
especially in the present context, is that it is naturally expressed in
terms of the Fourier space density fields which are measured by $21\,$cm
experiments.
A disadvantage of the Eulerian schemes is the increased difficulty of
including redshift-space distortions.
In this appendix we consider the impact of missing modes upon
reconstruction in the Eulerian scheme to build intuition about their impact.
We restrict ourselves to real-space measures, set the bias to $1$ and
the noise to zero for modes which have measurements.  This preserves the
main features of the problem while simplifying the presentation.

Heuristically, reconstruction works by ``undoing''  some of the nonlinear
time evolution which decreases and shifts the BAO peak.  For Eulerian
reconstruction, \citet{Sch15} approximate the initial density field by
Taylor expanding the observed field at late times, specifically subtracting
$\Delta t\, \dot{\delta}$.
Using the continuity equation to express $\dot{\delta}$ in terms of the
divergence of $(1+\delta)v$, and with an appropriate choice of $\Delta t$
some algebra gives \citep{Sch15}
\begin{equation}
  \delta_r(\bx) = \delta_{\rm obs}(\bx)
  - \bs(\bx)\cdot\vec{\nabla}\delta_{\rm obs}(\bx)
  - \delta_{\rm obs}(\bx)\delta_R(\bx) \; .
\end{equation}
Here ${\bs}(\bk)=-i\bk/k^2\ \delta_{\rm obs}(\bk)W(k)$
is the shift vector and $\delta_R(\bk)=\delta_{\rm obs}(\bk)W(k)$,
with a smoothing kernel $W(k)$ (usually a Gaussian).
In Fourier space, this becomes
\begin{equation}
  \delta_r(\bk) = \delta_{\rm obs}(\bk) -
  \int\frac{d^3p}{(2\pi)^3} \mathcal{K}(\bk-\mathbf{p},\mathbf{p})
  \delta_{\rm obs}(\bk-\mathbf{p})\delta_{\rm obs}(\mathbf{p})
\end{equation}
where by changing the kernel, $\mathcal{K}$, we can now describe more
general reconstruction methods.

The reconstructed power spectrum is
$P_{\rm rec}(k)\propto \langle \left| \delta_r^2\right|\rangle$
as before.  The leading contribution to enhancing the BAO features,
i.e.,~to reconstruction, was found by \citet{Sch15} to be (their Eq.~69): 
\begin{equation}
  \left\langle\Delta^{(3)}(\bk)\delta_L(\bk')\right\rangle =
  3(2\pi)^3\delta(\bk+\bk') P_{\rm lin}(k)
  \int\frac{d^3k_1}{(2\pi)^3} D_3({\bf k}_1,-{\bf k}_1,{\bf
    k})P_L(k_1) \; .
\label{eqn:d3def}
\end{equation}
This 3-point term can be seen to explicitly enhance the oscillations over
those in $P_{\rm obs}(k)$, restoring some of the signal that is damped by
non-linear evolution and bringing the result closer to $P_L(k)$.

In more detail,
\begin{equation}
  D_3({\bf k_1,k_2,k_3}) = -\frac{2}{3}[\mathcal{K}(\bk_1,\bk_2+\bk_3)
  F_2(\bk_2,\bk_3) + \; {\rm cyclic} ] \; ,
\label{eqn:d3term}
\end{equation}
where
\begin{equation}
  F_2(\bk_1,\bk_2)  = \frac{5}{7} +
  \frac{1}{2}\left(\frac{k_1}{k_2} +\frac{k_2}{k_1}\right)
  \hat{k}_1 \cdot \hat{k}_2 +
  \frac{2}{7} \left(\hat{k}_1 \cdot \hat{k}_2\right)^2 \; .
\end{equation}
is the usual perturbation theory kernel.
For the \citet{Sch15} EGS reconstruction method
\begin{equation}
  {\mathcal K}({\bf k}_1,{\bf k}_2) = \frac{1}{2} \left[W(k_1) +
  W(k_2)\right]+ W(k_1) \frac{\bk_1 \cdot \bk_2}{k_1^2} \; ,
\end{equation}
and, defining $\mu=\hat{k}_1\cdot\hat{k}$, 
\begin{eqnarray}
  D_3({\bf k_1,-k_1,k})&=&
  \frac{2}{3}\left[-\frac{1}{2}(W(\bk+\bk_1)+W(\bk-\bk_1)) \right.
  \left[\frac{5}{7} + \frac{2}{7} \mu^2\right] \nonumber \\
  &&-\frac{1}{4} \left[ W(\bk+\bk_1)-W(\bk-\bk_1)\right]\mu
  \left[\frac{k}{k_1} + \frac{k_1}{k}\right] \nonumber \\
  && \left. + W(k_1)\left[\frac{5}{7} +
  \mu^2\left(\frac{9}{7} + \frac{k^2}{k_1^2}\right)\right]\right]
  \; .
\label{eqn:d3exp}
\end{eqnarray}

\begin{figure}
\begin{center}
\resizebox{\columnwidth}{!}{\includegraphics{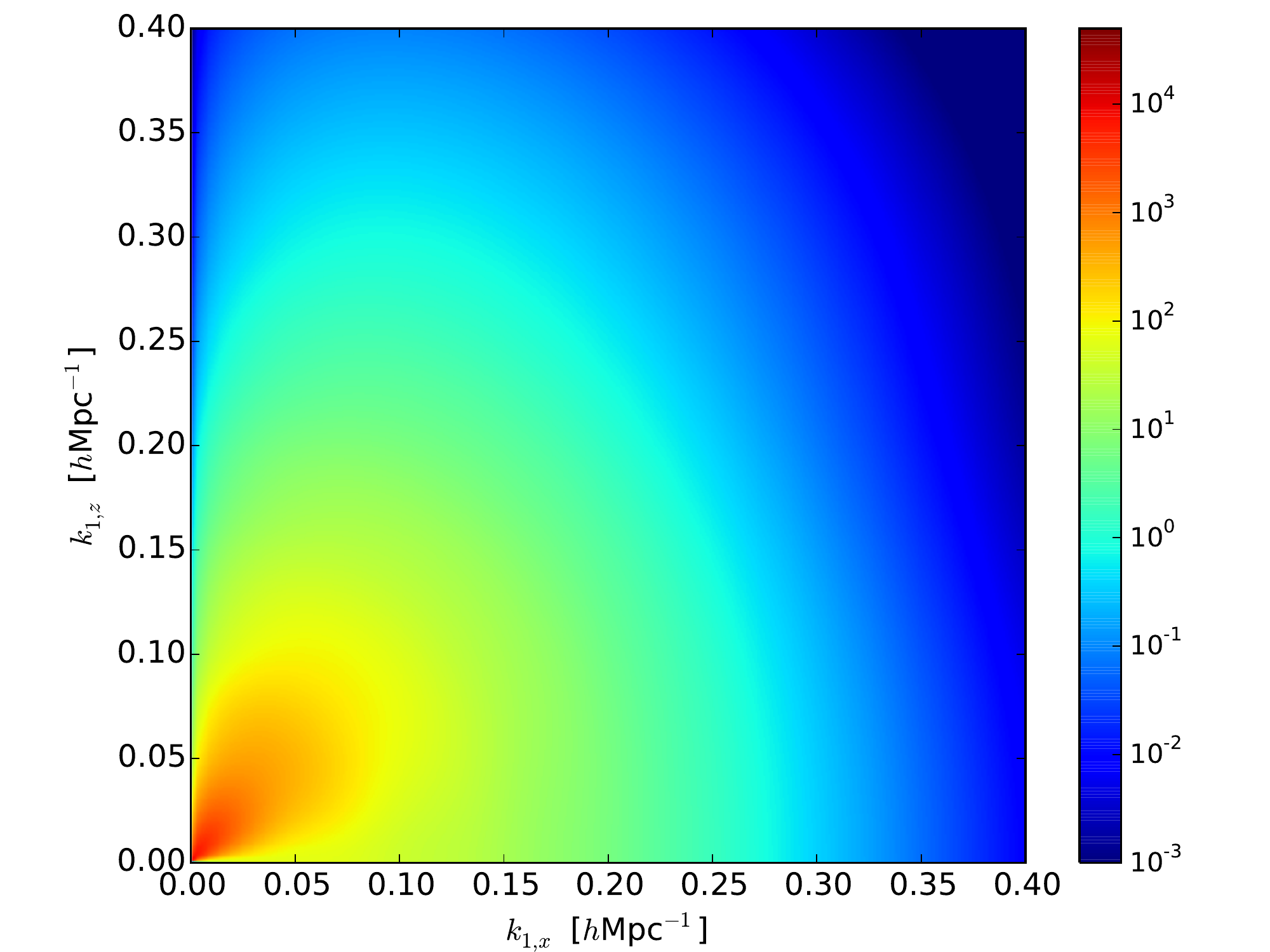}}
\resizebox{\columnwidth}{!}{\includegraphics{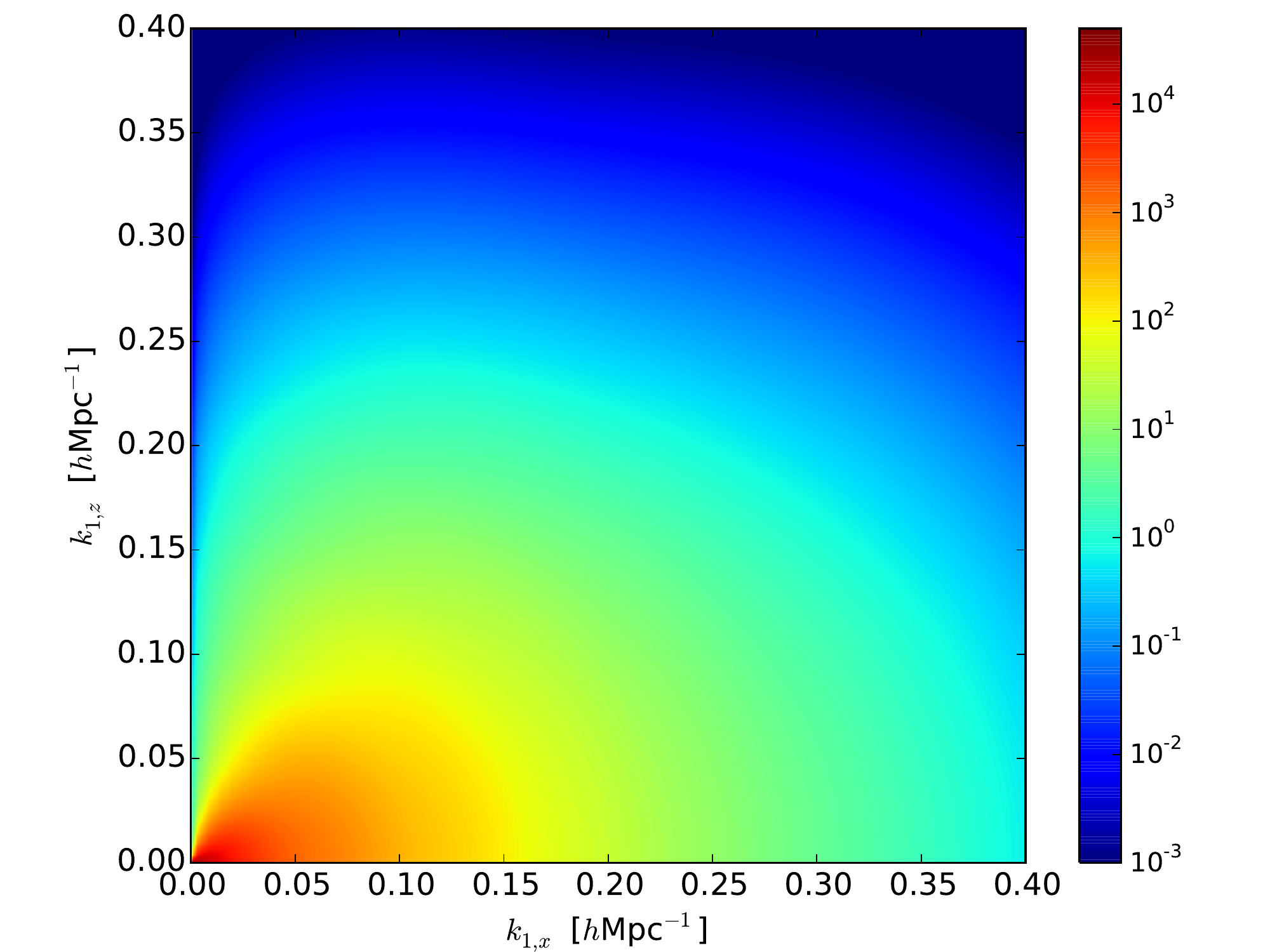}}
\end{center} 
\caption{Two slices in the $\hat{k}_{1,x},\hat{k}_{1,z}$ plane of the
integrand $D_3({\bf k_1,-k_1,k}) P_L(k_1)k_{1,\perp}$ in Eq.~\ref{eqn:d3def}.
Its integral is the dominant contribution in reconstruction of the linear
power spectrum $P_L({\bf k})$, as noted in \citet{Sch15}.
The top figure has the contribution for observed mode
$\bk=0.15\hat{k}_\parallel=0.15\hat{k}_z$,
and the bottom figure corresponds to $\bk=0.15\hat{k}_x$.
The measure of integration is
$d k_\parallel dk_\perp = dk_\parallel d k_x$
in this plane, thus area is a measure of the size of the contribution to
reconstruction.  Note that the color scale is logarithmic in this plot,
unlike the earlier plots.}
\label{fig:d3kperpP}
\end{figure}

Fig.~\ref{fig:d3kperpP} shows the contribution of different modes to
$D_3(\bk_1,-\bk_1,\bk)P_L(k_1)k_{1,\perp}$.  The modes which contribute
most to $D_3$ contribute the most to restoring the BAO signal, i.e.,~to
reconstruction.  Two different modes with $k=0.15\,h\,{\rm Mpc}^{-1}$
are shown, along the $\hat{k}_{1,z}$ and $\hat{k}_{1,x}$ axes
(the latter shown in the $\hat{k}_{1,z} -\hat{k}_{1,x}$ plane).
Equal area in this plot gets equal weight in the integral, so that
the relative importance of different modes can be read off more easily.
\footnote{For $\bk \propto \hat{k}_z$ the $\bk_1$ modes with large contributions seem closer to the
$\hat{k}_{1,z}$
axis than for $\bk \propto \hat{k}_{1,x}$.
We thank the
anonymous referee for pointing this out.
}

As before, if an interferometer does not measure a given $\bk_1$ mode, it
will not contribute to $D_3$ but it will contribute to the damping of the
signal in $P_{\rm obs}$.
A simple way to see the effect of missing modes to reconstruction of
a mode $\bk$ is to ask what impact removing contributions from $\bk_1$
and $\bk\pm\bk_1$ would have.  Visually it is clear that the modes where
the color scale is red or orange contribute the majority of $D_3$, so
losing these modes has the largest detrimental effect on reconstruction.
This agrees with our the intuition obtained from the Lagrangian theory
exposition in the main text.

\end{appendix}

\label{lastpage}
\end{document}